\documentclass[aps,pra,twocolumn,superscriptaddress]{revtex4-2}
\usepackage{amsfonts}
\usepackage{amsmath}
\usepackage{mathrsfs}
\usepackage{amssymb}
\usepackage{graphicx}
\usepackage{float} 
\usepackage{dcolumn}
\usepackage{bm}
\usepackage{color}
\usepackage[colorlinks=true,linkcolor=blue,anchorcolor=blue,citecolor=blue,urlcolor=blue]{hyperref}

\begin{document}

\title{Spin-orbit-coupling-induced phase separation in trapped Bose gases}

\author{Zhiqian Gui}
\affiliation{Department of Physics, Shanghai University, Shanghai 200444, China}

\author{Zhenming Zhang}
\affiliation{CAS Key Laboratory of Quantum Information, University of Science and Technology of China, Hefei 230026, China}

\author{Jin Su}
\affiliation{Department of Basic Medicine, Changzhi Medical College, Changzhi 046000, China}

\author{Hao Lyu}
\affiliation{Quantum Systems Unit, Okinawa Institute of Science and Technology Graduate University, Onna, Okinawa 904-0495, Japan}

\author{Yongping Zhang}
\email{yongping11@t.shu.edu.cn}
\affiliation{Department of Physics, Shanghai University, Shanghai 200444, China}

\begin{abstract}
In a trapped spin-$1/2$ Bose-Einstein condensate with miscible interactions, a two-dimensional spin-orbit coupling can introduce an unconventional spatial separation between the two components. We reveal the physical mechanism of such a spin-orbit-coupling-induced phase separation.  Detailed features of the phase separation are identified in a trapped Bose-Einstein condensate.  
We further analyze differences of phase separation in Rashba and anisotropic spin-orbit-coupled Bose gases. 
An adiabatic splitting dynamics is proposed as an application of the phase separation.

\end{abstract}

\maketitle

\section{Introduction}
\label{introduction}

Phase separation is a generic phenomenon from classical physics to quantum physics, for example, 
the oil-water separation and spin Hall effect~\cite{Sinova}. Two-component atomic Bose-Einstein condensates (BECs) provide a tunable platform for the investigations of phase separation~\cite{Shin,Papp,Thalhammer,Tojo,Nicklas,Ota,HeL}. 
The two components can be realized by using different atomic species or same species with different electric hyperfine states.
Such a system features intra- and inter-component interactions. When the inter-component interactions dominate over the intra-component interactions, two components prefer to be phase-separated to minimize the inter-component interactions~\cite{Trippenbach,Petrov2015}.  The interactions for phase separation are called immiscible.  The immiscibility of two-component BECs are completely tunable in experiments. 
Phase separation effect induces rich physics in quantum gases, such as the formation of vector solitons and vortex-soliton structures, 
coherent spin dynamics, and pattern formations~\cite{Shrestha,Law,LeeKL,Roy,Eto,Bernier,Shirley,Mistakidis,Lous}.

In a two-component BEC, an artificial  spin-orbit coupling can be synthesized between different hyperfine states via Raman lasers~\cite{Lin,Goldman,Zhai2015,Zhang2016}. Such a Raman-induced spin-orbit coupling is one dimensional. Rashba spin-orbit coupling, which is two dimensional, has also been experimentally realized in BECs~\cite{WuZ,Valdes}. 
The implementation of spin-orbit coupling in BECs gives rise to exotic quantum phases and rich superfluid properties, 
which opens an avenue for simulating topological matters and exploring superfluid dynamics~\cite{LiY,Manchon,LiJR,Khamehchi,Kartashov,Hasan,Valdes,Frolian}. In a Raman-type spin-orbit-coupled BEC, a stripe phase~\cite{Wang} can exist in miscible interactions~\cite{LiY,LiJR}. In contrast, for a Rashba spin-orbit-coupled BEC, the stripe phase may appear in the immiscible regime~\cite{Ho2011}. 
Very interestingly, the authors of Refs.~\cite{Hu2012,Zhang2012} numerically found a spatially phase-separated ground state in a Rashba-coupled and harmonically trapped BEC with miscible interactions.
Such a ground state in the miscible regime is unexpected for the usual two-component BEC without spin-orbit coupling. 
the authors of Ref.~\cite{Hu2012} identified that the exotic phase separation satisfies a combined symmetry of parity and a spin flip. 
The existence of this state is attributed to the authors of Refs.~\cite{Zhang2012,Song2013} to a spin-dependent force.  The force is intrinsic in the presence of  Rashba spin-orbit coupling and drives the two components moving in opposite directions.  The force concept provides an intuitive picture for the unexpected phase separation. However, its weakness is obvious. The force is proportional to the square of the Rashba spin-orbit coupling strength. Therefore, a large strength is expected to generate a larger spatial separation. In contrast, the numerical results show that the separation decreases with an increasing strength~\cite{Zhang2012,Song2013}.   
So far, the physical origin of the unconventional phase separation in the miscible regime is yet to be addressed.  It calls for an unambiguous interpretation, since the phase separation has already found a broad application in other excited states.  In Refs.~\cite{Sakaguchi,Xu2015}, a spin-orbit-coupled bright soliton was found to be spatially separated in the center of mass between the two components. The dynamics of the separation in bright solitons is analyzed by varying spin-orbit coupling strength~\cite{Wang2020}.  A spin-orbit-coupled single-vortex state, in which each component carries a singly quantized vortex, shows spatial separation between two components, and the separation is inversely proportional to spin-orbit coupling strength~\cite{Kato2017}. Recently, dynamics of the separation is triggered by a sudden quench of spin-orbit coupling strength in a trapped BEC~\cite{Ravisankar2021}.
All mentioned separations occur in the miscible regime, causing a counterintuitive expectation.

In this paper, we provide the physical mechanism for the unconventionally spin-orbit-coupling-induced phase separation.  Eigenstates of a two-dimensional spin-orbit coupling have a momentum-dependent relative phase $\varphi(\vec{k})$ between the two components. Closely around a fixed momentum $\vec{k}_0$, the relative phase may present a linear dependence $\varphi(\vec{k})\propto (\vec{k}-\vec{k}_0)\cdot \vec{r}_0$ with a constant $\vec{r}_0$.  The linear dependence is a momentum kick to move two components relatively. The superposition of these eigenstates distributing around $\vec{k}_0$ constitutes a spatially separated wave packet.  The separation, whose amplitude can be calculated,  is a completely single-particle effect of spin-orbit coupling.  A weakly trapped BEC with two-dimensional spin-orbit coupling is a perfect platform to simulate the phase separation. The miscible interactions force atoms to condense at a certain spin-orbit-coupled momentum state with a momentum-dependent relative phase. Meanwhile, weak traps broaden the condensed momentum so that the condensation occupies momentum states in a narrow regime, which give rise to the linear dependence of the relative phase.  We numerically identify detailed features of spin-orbit-coupling-induced phase separation in a trapped BEC with miscible interactions by analyzing its ground states.  The phase separation matches with the single-particle prediction when spin-orbit coupling strength dominates. We also compare the separation differences between Rashba and an anisotropic spin-orbit coupling. Finally, as an application of the phase separation, we propose an adiabatic splitting dynamics.

The paper is organized as follows. In Sec.~\ref{mechanism}, the physical mechanism of spin-orbit-coupling-induced phase separation is unveiled.  The separation amplitudes are predicted.  From the mechanism, we know that the separation is a single-particle effect.  In Sec.~\ref{Rashba}, we identify the separated features of ground states in a trapped BEC with Rashba spin-orbit coupling by the imaginary-time evolution method and the variational method.   In Sec.~\ref{AnistropicSOC}, we reveal the effect of the anisotropy of spin-orbit coupling on the phase separation.  In Sec.~\ref{Adiabaticdynamics}, we propose an adiabatic dynamics to dynamically split two components basing on the phase separation.  For the completeness of our discussion, immiscible-interaction-induced phase separation is shown in Sec.~\ref{Immiscibilty}.
The conclusion follows in Sec.~\ref{conclusion}.

\section{Spin-orbit-coupling-induced phase separation}
\label{mechanism}

Rashba spin-orbit-coupling-induced phase separation is a completely single-particle effect. We reveal the physical origin of such phase separation.  The Rashba spin-orbit-coupled Hamiltonian is
\begin{equation}
\label{Hsoc}
H_\text{SOC}=\frac{p_x^2+p^2_y}{2}+\lambda (p_{x} \sigma_{y}- p_{y} \sigma_{x}),
\end{equation}
where $p_x$ and $p_y$ are the momenta along the $x$ and $y$ directions respectively, $\lambda$ is the spin-orbit coupling strength, and $\sigma_{x,y}$ are spin-1/2 Pauli matrices.  The eigenenergy of the Hamiltonian has two bands. The lower band is 
\begin{equation}\label{Band}
E = \frac{k_x^2 + k_y^2}{2} - \lambda\sqrt{k_x^2 + k_y^2},
\end{equation}
with associated eigenstates being
\begin{equation}\label{Eigenstate}
\Phi = \frac{1}{\sqrt{2}}e^{ik_x x+ik_y y}
\begin{pmatrix}
e^{i\frac{\varphi}{2}} \\ e^{-i\frac{\varphi}{2}}
\end{pmatrix}.
\end{equation}	
Since the Hamiltonian possesses continuously translational symmetry, the eigenstates are plane waves with  $k_{x,y}$ being the quasimomenta along the $x$ and $y$ directions, respectively.  The outstanding feature is that Rashba spin-orbit coupling generates a relative phase $\varphi$ between the two components, which satisfies 
\begin{equation}\label{phi}
\tan(\varphi) = \frac{k_x}{k_y}.
\end{equation}
It is noted that $(k_x,k_y)=(0,0)$ is a singularity, closely around which the relative phase cannot be defined. Therefore, the eigenstate in Eq.~(\ref{Eigenstate}) works beyond the regime around the singularity. 

We construct a wave packet by superposing these eigenstates,
\begin{equation}\label{Packet}
\Psi = \int_{-\infty}^{\infty} dk_x dk_y \mathcal{G}({\mathbf{k}}-\mathbf{\bar{k}} )\Phi,
\end{equation}
with the superposition coefficient $\mathcal{G}$ being a momentum-dependent localized function centering around $\mathbf{\bar{k}}$. For a straightforward illustration, we take a Gaussian distribution as an example,
\begin{equation}\label{Gaussian}
\mathcal{G} ({\mathbf{k}}- \mathbf{\bar{k}}) = \frac{1}{2\pi \sqrt{\Delta_x\Delta_y}} e^{-\frac{(k_x - \bar k_x)^2}{2\Delta_x} -\frac{(k_y - \bar k_y)^2}{2\Delta_y}}.
\end{equation}		
The Gaussian distributed superposition coefficient is centered at $\mathbf{\bar{k}}=(\bar k_x,\bar k_y)$  with the packet widths $\sqrt{\Delta_{x,y}}$ along the $x$ and $y$ directions. If the widths are narrow, the superposition mainly happens around $\mathbf{\bar{k}}$. Therefore, we analyze the eigenstates around $\mathbf{\bar{k}}$, and the relative phase becomes
\begin{equation}\label{Phase}
\varphi({\mathbf{k}} ) \approx \varphi(\mathbf{\bar{k}}) + \left.  (k_x - \bar k_x)\frac{\partial \varphi}{\partial k_x}\right |  _{\mathbf{\bar{k}}} + \left.  (k_y - \bar k_y)\frac{\partial \varphi}{\partial k_y}\right |_{\mathbf{\bar{k}}},
\end{equation}	
which is linearly dependent on the momenta $k_{x,y}$. This is true since expanding any continuous function around a certain 
parameter point leads to dominant linear-dependence.  Such linear dependence in Eq.~(\ref{Phase}) induced a momentum kick, generating the relative motion between the two components. 
After substituting the Gaussian distribution in Eq.~(\ref{Gaussian}) and $\varphi$ in Eq.~(\ref{Phase}) into Eq.~(\ref{Packet}) and performing integration, we obtain the wave packet,
\begin{align}\label{wavefunction}
\Psi = &\frac{1}{\sqrt{2}} e^{i\bar{k}_x x+i\bar{k}_y y } \notag \\
&\times
\begin{pmatrix}
e^{-\frac{\Delta_x}{2}\left [ x+ \frac{1}{2}\frac{\partial \varphi (\mathbf{\bar{k}} )}{\partial k_x}\right ]^2 -\frac{\Delta_y}{2} \left[ y+\frac{1}{2}\frac{\partial \varphi (\mathbf{\bar{k}})}{\partial k_y} \right]^2 + i\frac{\varphi(\mathbf{\bar{k}} )}{2}} \\
e^{-\frac{\Delta_x}{2}\left [ x- \frac{1}{2}\frac{\partial \varphi (\mathbf{\bar{k}})}{\partial k_x}\right ]^2 -\frac{\Delta_y}{2} \left[ y-\frac{1}{2}\frac{\partial \varphi (\mathbf{\bar{k}} )}{\partial k_y} \right]^2 - i\frac{\varphi(\mathbf{\bar{k}} )}{2}}
\end{pmatrix}.
\end{align}
The outstanding feature of the resultant wave packet is that the two components have a relative position displacement.  The displacements along the $x$ and $y$ directions are
 \begin{equation}
 \left.\frac{\partial \varphi(\mathbf{k} )}{ \partial k_x}\right|_{\mathbf{\bar{k}}}= \frac{\bar{k}_y}{ \bar{k}_x^2+\bar{k}_y^2},\ \ 
\left. \frac{\partial \varphi(\mathbf{k} )}{ \partial k_y}\right|_{\mathbf{\bar{k}}}= -\frac{\bar{k}_x}{ \bar{k}_x^2+\bar{k}_y^2}.
 \label{Shift}
 \end{equation}
The nonzero displacements give rise to a phase separation between two components. From the construction of the phase-separated wave packets, 
we can see that the origin of the phase separation is the existence of the momentum-dependent relative phase in eigenstates and the occupation of these eigenstates confined in a narrow momentum regime. 

The Rashba spin-orbit coupled BEC is an ideal platform to generate such a phase-separated state. The lower band in Eq.~(\ref{Band}) has  infinite energy minima which locate at the quasimomenta satisfying $k_x^2+k_y^2=\lambda^2$; therefore, $k_x=\lambda \cos(\theta)$ and $k_y=\lambda \sin(\theta)$ with $\theta $ being an angle. The interacting atoms spontaneously choose one of the energy minima to condense and form a BEC~\cite{Wang, Hu2012, Zhang2012}.
This means that $\theta$ is spontaneously chosen to be a value $\bar{\theta}$.  In real atomic BEC experiments, traps are inevitable. A weak harmonic trap naturally broadens the BEC momentum giving rise to a Gaussian distribution centered at $(\bar{k}_x,\bar{k}_y)=\lambda(\cos(\bar{\theta}), \sin(\bar{\theta}))$. Furthermore, the broadening is narrow so that Eq.~(\ref{Phase}) is satisfied. Consequently, the Rashba-coupled BEC  presents as a phase separated state in Eq.~(\ref{wavefunction}) with $\partial \varphi (\mathbf{k} ) / \partial k_x|_{\mathbf{\bar{k}}}=\sin(\bar{\theta})/\lambda$ and $\partial \varphi (\mathbf{k} )| / \partial k_y|_{\mathbf{\bar{k}}}=-\cos(\bar{\theta})/\lambda$.  The position displacement is inversely proportional to the spin-orbit coupling strength $\lambda$, which clearly indicates Rashba spin-orbit-coupling-induced phase separation. When the strength goes to zero ($\lambda \approx 0$), the momentum $(\bar{k}_x, \bar{k}_y)\approx (0,0)$ becomes a singularity so that the eigenstate in Eq.~(\ref{Eigenstate}) is not physical. Without the spin-orbit coupling, the BEC becomes the conventional one, and there is no phase separation between two components. 
If the strength is enhanced gradually from zero, the position displacements should continuously increase from zero to catch up with the predicted value  $[\sin(\bar{\theta})/\lambda, -\cos(\bar{\theta})/\lambda]$. When the strength $\lambda$ is large enough, the position displacements decrease towards zero again since they are inversely proportional to $\lambda$.   
In this case, the plane-wave phase $(\bar{k}_xx+\bar{k}_yy)$ dominates, while the relative phase in the eigenstates [Eq.~(\ref{Eigenstate})] is independent of $\lambda$, 
i.e., $\tan(\varphi)=\bar{k}_x/\bar{k}_y=\cot(\bar{\theta})$.
Consequently, the effect of the relative phase is obliterated by the plane-wave phase, and phase separation disappears.

According to the above mechanism of the phase separation, if there is no weak trap to broaden the condensed momentum, the spin-orbit-coupled BEC can not present the position displacement.  This is why a spatially homogeneous BEC with spin-orbit coupling does not show phase separation as studied in most literature.  Nevertheless, to broaden the condensed momentum without traps, we may consider spatially localized excitation states, such as bright solitons and vortices.  These spatially self-trapped states naturally broadens the condensed momentum. Therefore, the resultant phase separation between two components in Rashba spin-orbit-coupled bright solitons and quantum vortices, which have been numerically revealed in Refs.~\cite{Sakaguchi,Xu2015,Kato2017}, can be understood by a generalization of our mechanism.  
Interestingly, the position displacement of quantum vortex is inversely proportional to the spin-orbit coupling strength as uncovered numerically in Ref.~\cite{Kato2017}, can be explained 
unambiguously.

We emphasize that the spin-orbit-coupling-induced phase separation only works for a two-dimensional spin-orbit coupling.  For an one-dimensional spin-orbit coupling, i.e., the Raman-induced one, the single-particle Hamiltonian is $H'=p_x^2/2 +\lambda p_x\sigma_z+\Omega \sigma_x$ with the $\Omega$ being the Rabi frequency due to Raman lasers~\cite{Lin, Zhang2016}. The lower energy band of this system is $E=k_x^2/2 - \sqrt{\lambda^2k_x^2+\Omega^2}$ with the eigenstates being $\Phi=e^{ik_xx}[-\sin(\Theta), \cos(\Theta)]^T$. Here, $\tan(\Theta)=\Omega/(\lambda k_x)$, and $T$ is the transpose operator. It is noted that there is no momentum-dependent relative phase in the eigenstates. Therefore, the Raman-induced spin-orbit coupling, in principle, can not generate the phase separation. 

In the above, we revealed the physical 
mechanism of Rashba spin-orbit-coupling-induced phase separation.  We demonstrate that a weakly trapped spin-orbit-coupled BEC satisfies the requirements of the mechanism. The quantum phase in trapped spin-orbit-coupled BECs may be phase separated states. The spin-orbit-coupling-induced phase separation is a single-particle effect. The role of nonlinearity in the BEC is to spontaneously choose one energy minimum for condensation. In the following, we study the ground states of a trapped spin-orbit-coupled BEC and identify features of spin-orbit-coupling-induced phase separation.

\section{Rashba spin-orbit-coupling-induced phase separation in trapped BECs}
\label{Rashba}

We consider a quasi-two-dimensional spin-1/2 BEC with Rashba spin-orbit coupling. The trap frequency $\omega_z$ along the $z$ direction is assumed to be very large so that the dynamics is completely frozen into the ground state of the $z$-directional harmonic trap. 
Such a strong trap can be implemented by an optical lattice in the $z$ direction in experiments. After integrating the atomic state along the $z$ direction, we are left with a quasi-two-dimensional system.  Rashba spin-orbit coupling can be artificially implemented by an optical Raman lattice~\cite{WuZ}, generating the Hamiltonian $H_\text{SOC}$ shown in Eq.~(\ref{Hsoc}). The spin-orbit-coupled BEC is described by the following Gross-Pitaevskii (GP) equation,
\begin{align}
i\frac{\partial\Psi}{\partial t}=\left( H_\text{SOC}+V+H_\text{int}\right) \Psi.
\end{align}
with  $\Psi=\left( \Psi_1,\Psi_2\right) ^T$ being the two-component wave function.
The harmonic trap in the $x$-$y$ plane is $V=\frac{1}{2}\omega^2(x^2+y^2)$ with $\omega$ the dimensionless trap frequency.
$H_\text{int}$ denotes the nonlinear interactions,
\begin{align}
H_\text{int}=\left(\begin{matrix}
g\left|\Psi_{1}\right|^{2}+g_{12}\left|\Psi_{2}\right|^{2} & 0 \\
0 & g_{12}\left|\Psi_{1}\right|^{2}+g\left|\Psi_{2}\right|^{2}
\end{matrix}\right),
\end{align}
The GP equation is dimensionless, and  the units of length, time,  momentum, and energy are chosen as  $l_z=\sqrt{\hbar/(m\omega_z)}$, $1/\omega_z$, $\hbar/l_z$, and $\hbar \omega_z$, respectively. With the units, the inter and intracomponent interaction coefficients become
 $g=Na\sqrt{8\pi}/l_z$ and $g_{12}=Na_{12}\sqrt{8\pi}/l_z$.
Here, $N$ is the atom number,  and $a$ and $a_{12}$ are the corresponding $s$-wave scattering lengths, respectively.
The wave functions satisfy the normalization condition, $\int dx dy (|\Psi_1|^2+|\Psi_2|^2)=1$. 
In numerical calculations,  experimentally accessible parameters are used.  The typical trap frequency is $\omega_z=2\pi\times200$ Hz, leading to the units of length and time $l_z=0.76\mu$m and $1/\omega_z=0.8$ms respectively. 
 $a\sim 100~a_0$ with  $a_0$ being the Bohr radius, and $N\sim300$, lead to $g\sim 10$.
The spin-orbit coupling strength can be changed by tuning the parameters of Raman lasers in the experiments~\cite{WuZ}.

When $g>g_{12}$, the interactions are miscible. We first study the ground states of the system in this regime by performing the imaginary-time evolution of the GP equation. 
The evolution is numerically implemented by the split-step Fourier method. The window
of two-dimensional space is chosen as $(x, y)\in [-6\pi,6\pi] $ and
is discretized into a 256 × 256 grid. 

\begin{figure}[t]
\centering
\includegraphics[width=0.98\linewidth]{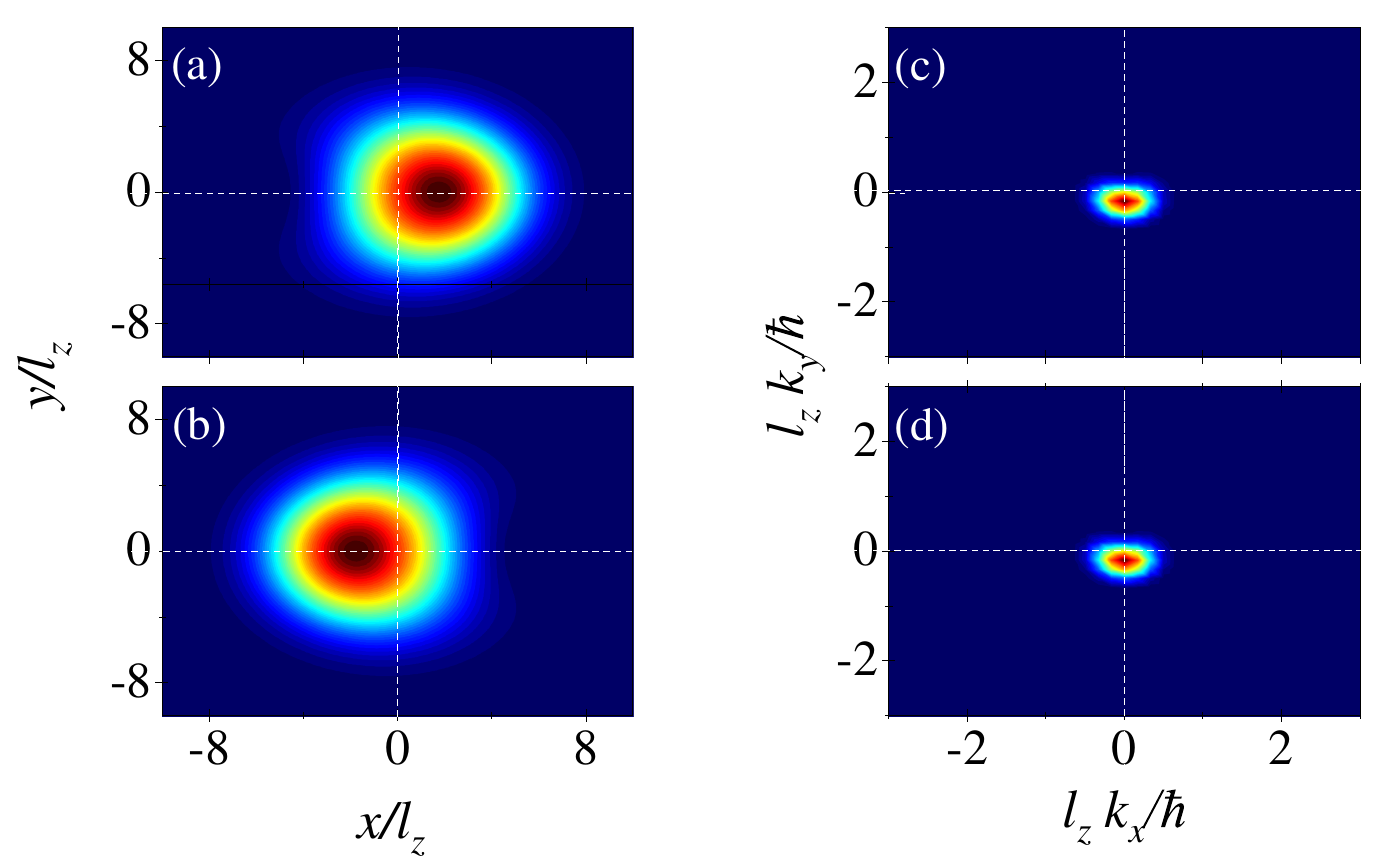}
\caption{Ground state of a trapped Rashba spin-orbit-coupled BEC with miscible interactions.
(a), (b) Density distributions $|\Psi_1|^2$ and $|\Psi_2|^2$ in the coordinate space. 
(c), (d) Density distributions $|\Psi_1|^2$ and $|\Psi_2|^2$ in the momentum space.
The parameters are $\omega/\omega_z=0.1$, $l_z\lambda/\hbar=0.2$, $g=12$ and $g_{12}=8$.}
\label{fig:figure1}
\end{figure}

A typical result is shown in Fig.~\ref{fig:figure1}.  As expected from the prediction in the previous section, the ground state is phase-separated. The two components are spatially separated along the $x$ direction, as shown by Figs.~\ref{fig:figure1}(a) and~\ref{fig:figure1}(b). The ground state spontaneously chooses $\bar{\theta}=-\pi/2$ so that the atoms condense at $(\bar{k}_x,\bar{k}_y)=(0,-\lambda)$, which can be seen from the momentum-space density distributions in Figs.~\ref{fig:figure1}(c) and~\ref{fig:figure1}(d).  In this case, according to Eq.~(\ref{Shift}), the position displacement occurs along the $x$ direction, and the first component shifts by $1/(2\lambda)$ on the right side and the second component shifts oppositely by $1/(2\lambda)$ on the left side.

%%%%%%%%%%%%%%%%%%%%%%%%%%%%%%%%%%%%%%%%%%%%%
%%%%%%%%%%%%%%%%%%%%%%%%%%%%%%%%%%%%%%%%%%%%%

%%%%%%%%%%%%%%%%%%%%%%%%%%%%%%%%%%%%%%%%%%%%%
\begin{figure}[t]
\centering
\includegraphics[width=0.98\linewidth]{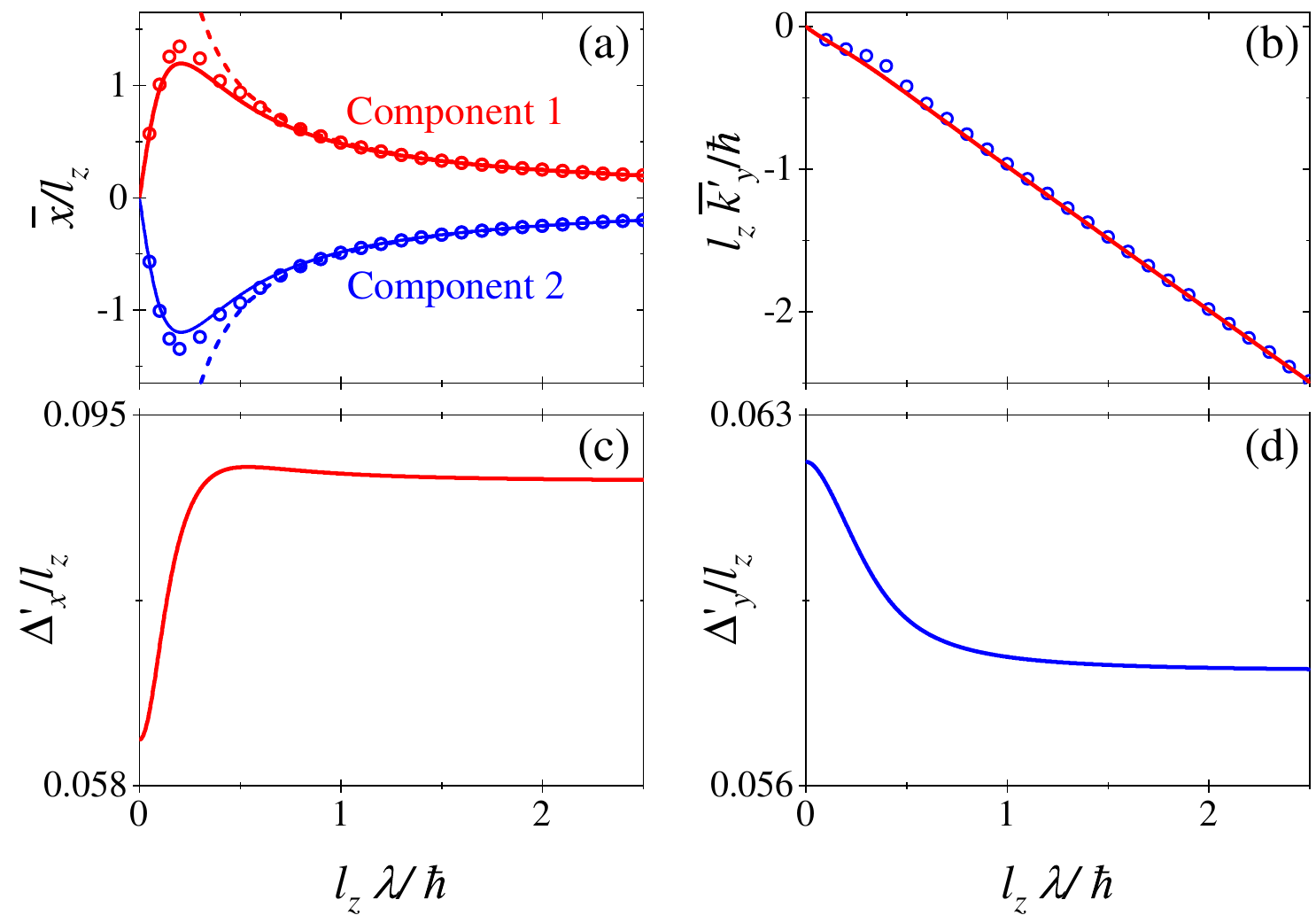}
\caption{ Rashba spin-orbit-coupling-induced phase separation in a trapped BEC. The parameters are $\omega/\omega_z=0.1$, $g=12$, and 
$g_{12}=8$.
(a) The center of mass for the two components along the $x$ direction as a function of the spin-orbit coupling strength $\lambda$.  The solid lines are from the variational method, and the circles are obtained by the imaginary-time evolution of the GP equation. The dashed lines are $\pm 1/(2\lambda)$ predicted from the single-particle model. The red (blue) color represents the first (second) component. (b) The condensed momentum $\bar{k}'_y$ as a function of $\lambda$. The red solid line and blue circles are obtained by the imaginary-time evolution and the variational method, respectively.
The variational parameters $\Delta'_{x,y}$ are shown in (c) and (d).}
\label{fig:separation}
\end{figure}
%%%%%%%%%%%%%%%%%%%%%%%%%%%%%%%%%%%%%%%%%%%%%

In the presence of interactions, it is impossible to construct an analytical wave function of the ground state from the procedure demonstrated in the previous section. Nevertheless, the single-particle wave functions in Eq.~(\ref{wavefunction}) and phase-separated results shown in Fig.~\ref{fig:figure1} stimulate us to use a trial wave function to study the phase separation by the variational method~\cite{Song2013}. The trial wave function is assumed to be
\begin{align}
\label{eq:ansatz}
\Psi(x,y)
=\frac{(\Delta'_x\Delta'_y)^{\frac{1}{4}}}{\sqrt{2\pi} }e^{i\bar{k}'_y y}
\left(\begin{array}{cc}
e^{-\frac{\Delta'_x}{2}(x-\delta_x)^2-\frac{\Delta'_y}{2}y^2} \\
e^{-\frac{\Delta'_x}{2}(x+\delta_x)^2-\frac{\Delta'_y}{2}y^2}
\end{array}\right).
\end{align}
Here, we assume that the atoms spontaneously condense at $(0,\bar{k}'_y)$ in momentum space and therefore the phase separation only happens along the $x$ direction with the relative position displacement $2\delta_x$. $1/\sqrt{\Delta'_{x,y}}$ characterize the widths of the wave packet along the $x,y$ directions.
The unknown parameters  $\bar{k}'_y, \delta_x, \Delta'_{x,y}$  are to be determined by minimizing the energy functional,
\begin{align}
\mathcal{E}&= \int dxdy\Psi^* (H_\text{SOC} +V) \Psi \notag \\
&\phantom{={}}+\int dxdy \left[\frac{g}{2}(|\Psi_1|^4+|\Psi_2|^4) + g_{12} |\Psi_1|^2|\Psi_2|^2\right],
\end{align}
Substituting the trial wave function into the energy functional $\mathcal{E}$ leads to
\begin{align}
\mathcal{E}&= \frac{\bar{k}_y^{'2}}{2}
+\frac{\Delta'_{x}+\Delta'_{y}}{4}\left(1+\frac{\omega^2}{\Delta'_x\Delta'_y} \right) +\frac{1}{2}\omega^2\delta_x^{2} \notag\\
&+ \lambda (\Delta'_x\delta_x-\bar{k}'_y)e^{-\Delta'_x \delta_x^2}
+\frac{\sqrt{\Delta'_{x}\Delta'_{y}}}{8 \pi}\left( g+g_{12} e^{-2\Delta'_{x}\delta_{x}^{2} }  \right) .
\label{eq:energy}
\end{align}
By minimizing the energy functional with respect to the unknown parameters, $\partial \mathcal{E}/\partial X=0$ $(X=\bar{k}'_y, \delta_x, \Delta'_{x,y})$, we obtain all information of the trial wave function. 
 The phase separation can be characterized by the center of mass of each component,
\begin{align}
\bar{\bm{r}}_{1,2}=\int \bm{r}|\Psi_{1,2}(\bm{r})|^2 d\bm{r},
\end{align}
with $\bm{r}=(x,y)$. In Fig.~\ref{fig:separation}(a), the solid lines show $\bar{x}_{1,2}=\pm \delta_x$ calculated from the variational method, 
while the results obtained by the imaginary-time evolution of the GP equation are demonstrated by the circles. 
We find that the results from the two calculation methods agree very well. Without spin-orbit coupling ($\lambda=0$), the conventional BEC has $\bar{x}_{1,2}=0$ and condensates at $\bar{k}'_y=0$, as shown in Fig.~\ref{fig:separation}(b). 
With the growth of $\lambda$, $\bar{k}'_y$ always increases linearly [see Fig.~\ref{fig:separation}(b)]. The displacement of $\bar{x}$ first increases drastically to a maximum value and then declines to the predicted $\pm 1/(2\lambda)$ obtained by the single-particle model [see the dashed lines in Fig.~\ref{fig:separation}(a)]. The dependence of the displacement on $\lambda$ exactly follows the expectation in the previous section.  
In the dramatic increase regime for $\bar{x}$, the variational parameters $\Delta'_{x,y}$ also change dramatically [see Figs.~\ref{fig:separation}(c) and~\ref{fig:separation}(d)].

%%%%%%%%%%%%%%%%%%%%%%%%%%%%%%%%%%%%%%%%%%%%%
\begin{figure}
\centering
\includegraphics[width=0.98\linewidth]{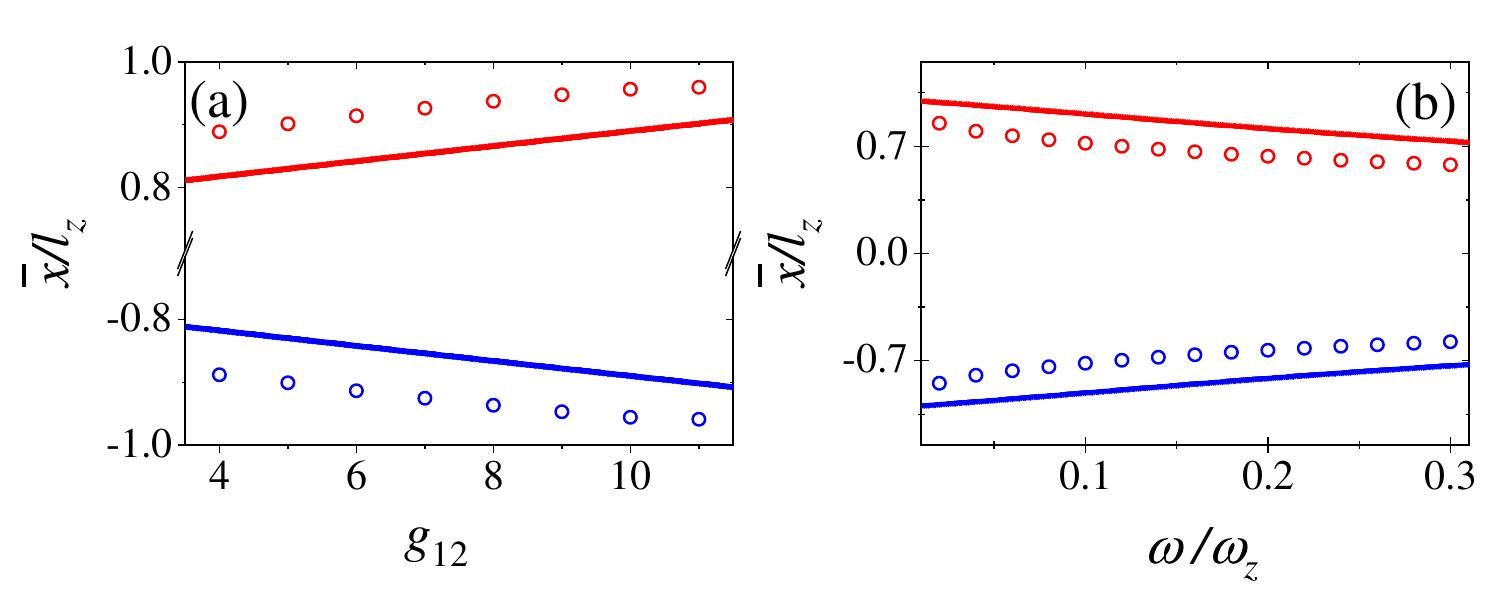}
\caption{
The center of mass of two components for a non-dominant $l_z\lambda/\hbar=0.5$. The solid lines are obtained by the variational method and the circles are from the imaginary-time evolution of the GP equation. The red and blue colors represent the first and second components, respectively.  (a) The center of mass as a function of the intercomponent interaction coefficient $g_{12}$. $\omega/\omega_z=0.1$ and $g=12$.  (b) The center of mass as a function of the trap frequency $\omega$. $g=12$ and $g_{12}=8$.}
\label{fig:interaction-trap}
\end{figure}
%%%%%%%%%%%%%%%%%%%%%%%%%%%%%%%%%%%%%%%%%%%%%

Rashba spin-orbit coupling introduces an intrinsic force, 
\begin{align}
\bm{F}=\frac{d \bm{p}}{dt}&=-\big[[\bm{r},H_\text{SOC}],H_\text{SOC}\big] \notag \\
&=2\lambda^2(\bm{p}\times\bm{e}_z)\sigma_z,
\end{align}
with $\bm{e}_z$ being the unit vector along the $z$ direction and $\bm{p}$ the atom momentum. The force originates from spin-orbit-coupling-induced anomalous velocity~\cite{MardonovPRL2015,MardonovPRA2015,MardonovPRA2018}.   Considering the ground states shown in Fig.~\ref{fig:separation}, the force operator in momentum space becomes $F_x=2\lambda^2\bar{k}'_y\sigma_z$ and $F_y=0$. 
The two components feel an opposite force $F_x$ along the $x$ direction. The ground states must compensate for the intrinsic force to reach equilibrium. It can be implemented by displacing two components opposite to the force. Since $\bar{k}'_y<0$ in the case shown in Fig.~\ref{fig:separation}, the first component is displaced towards to the right side and the second to the left side.
The force concept has been used in Refs.~\cite{Zhang2012,Song2013} to explain the phase separation. Since the force is proportional to $\lambda^2$,
it seems that a large displacement would be induced for a large $\lambda$. However, as shown in Fig.~\ref{fig:separation}(a), the dependence of the displacement on $\lambda$ does not follow the force. 
We can see that the intrinsic force cannot explain the phase separation in the large $\lambda$ regime.

Figure~\ref{fig:separation}(a) shows that the separation follows the single-particle prediction $\pm 1/(2\lambda)$ when $\lambda$ dominates. 
When $\lambda$ is weak, the displacement also depends on other parameters, such as nonlinear coefficients and the harmonic trap. In Fig.~\ref{fig:interaction-trap}(a), we plot the displacement $\bar{x}$ as a function of the inter-component interaction coefficient $g_{12}$ for a non-dominant $\lambda$. The displacement slightly rises with an increasing $g_{12}$, and it reaches the maximum when $g_{12}=g$. If $g_{12}>g$, the interactions become immiscible, leading to ground states different from the trial wave function in Eq.~(\ref{eq:ansatz}). The dependence of the displacement on the trap frequency is shown in Fig.~\ref{fig:interaction-trap}(b). We find that the displacement decreases
as the trap frequency increases. This is because the displacement requires more kinetic energy in a tight trap. 
It is noticed that there is a slight mismatching between the results from the variational method (the solid line) and the imaginary-time evolution (the circles) in Fig.~\ref{fig:interaction-trap}. The origin of such mismatching is that the Gaussian profile in the trial wave function in Eq.~(\ref{eq:ansatz}) cannot exactly describe the imaginary-time-evolution-generated wave function as shown in Fig.~\ref{fig:figure1}.

%%%%%%%%%%%%%%%%%%%%%%%%%%%%%%%%%%%%%%%%%%%%%
\begin{figure*}
\centering
\includegraphics[width=0.9\linewidth]{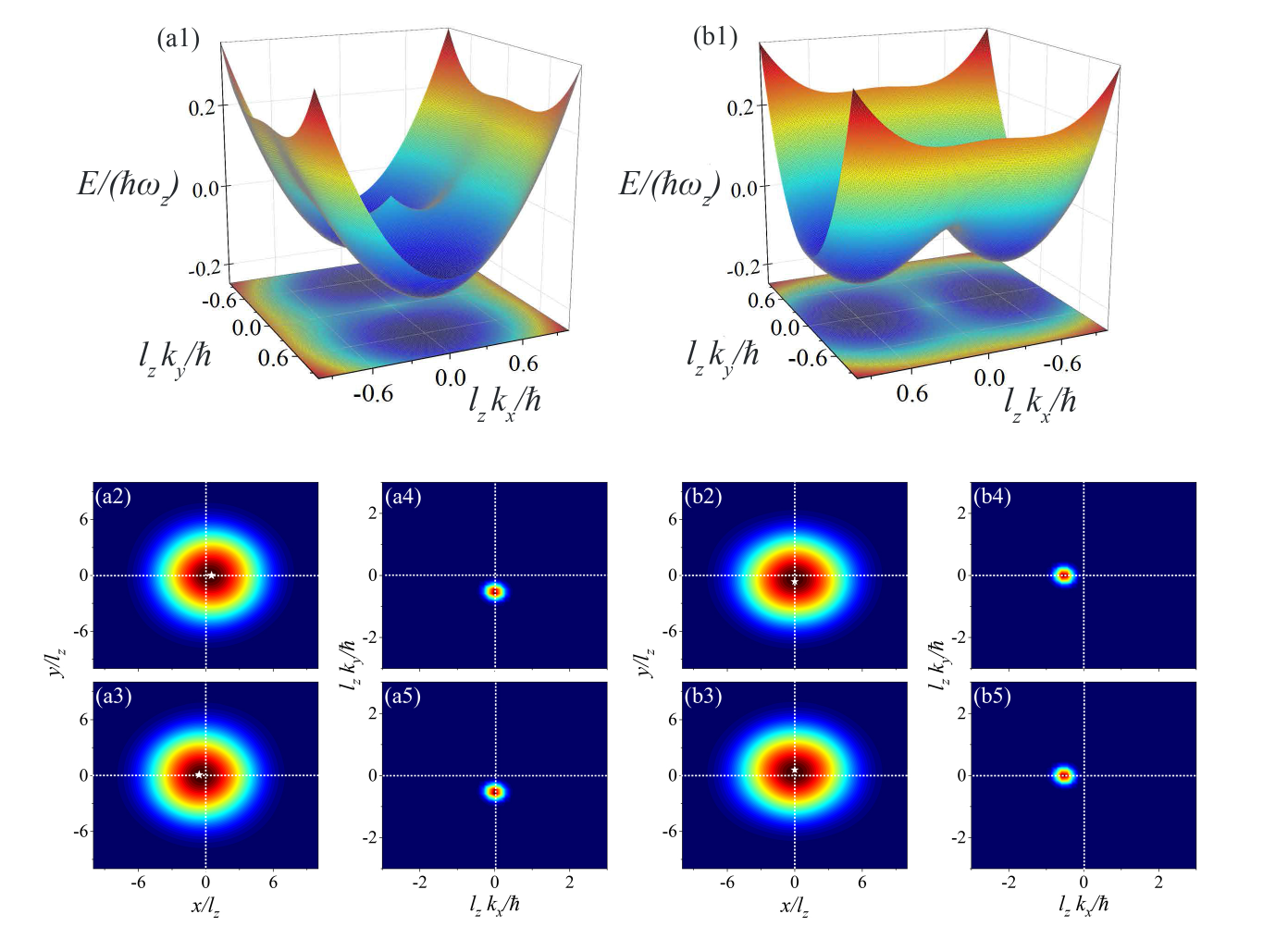}
\caption{ The anisotropic spin-orbit-coupling-induced phase separation in a trapped BEC. The parameters are $\omega/\omega_z=0.1$, $g=12$, and $g_{12}=8$. 
(a1) The lower band of $H'_\text{SOC}$ in Eq.~(\ref{Asoc}) with $l_z\lambda_1/\hbar=0.3$ and $l_z\lambda_2/\hbar=0.6$.
(a2) and (a4) show corresponding ground-state density distributions of  the first component $|\Psi_1|^2$ in coordinate and momentum spaces, respectively. (a3) and (a5) are for the second component $|\Psi_2|^2$. 
(b1)--(b5) are same as (a1)--(a5) but with $l_z\lambda_1/\hbar=0.6$ and $l_z\lambda_2/\hbar=0.3$.
In (a2), (a3), (b2), and (b3), white stars represent the center of wave packets predicted by the single-particle model.
}
\label{fig:anisotropy}
\end{figure*}
%%%%%%%%%%%%%%%%%%%%%%%%%%%%%%%%%%%%%%%%%%%%%

\section{The anisotropic spin-orbit-coupling-induced phase separation in trapped BECs}
\label{AnistropicSOC}

Rashba-spin-orbit-coupling-induced phase separation was analyzed in the previous section.  In the two-dimensional spin-orbit-coupled BEC experiment~\cite{WuZ}, the spin-orbit coupling strengths are tunable, which leads to an anisotropic coupling. 
It was revealed  that the anisotropic spin-orbit coupling has a great impact on ground states of a spatially homogeneous BEC~\cite{Ozawa}.
In this section, we study anisotropic-spin-orbit-coupling-induced phase separation.
The single-particle Hamiltonian of the anisotropic spin-orbit coupling is
\begin{equation}
\label{Asoc}
H'_\text{SOC}=\frac{p_x^2+p^2_y}{2}+\lambda_1 p_{x} \sigma_{y}-\lambda_2 p_{y} \sigma_{x},
\end{equation}
with the anisotropic strengths $\lambda_1 \ne \lambda_2$. 
The lower band of $H'_\text{SOC}$ is
\begin{align}
E= \frac{k_x^2 + k_y^2}{2} - \sqrt{\lambda_1^2 k_x^2 + \lambda_2^2 	k_y^2}.
\end{align}
with the associated eigenstates being the same as Eq.~(\ref{Eigenstate}) but having a different relative phase which can be written as
\begin{equation}\label{Aphi}
\tan(\varphi) = \frac{\lambda_1k_x}{\lambda_2k_y}.
\end{equation}
According to the mechanism of the spin-orbit-coupling-induced phase separation, the anisotropic coupling can generate position displacements related to the derivatives of the relative phase.  The displacements along the $x$ and $y$ directions are
 \begin{align}
&\left.\frac{\partial \varphi(\mathbf{k} )}{ \partial k_x}\right|_{\mathbf{\bar{k}}}= \frac{\lambda_1\lambda_2\bar{k}_y}{ \lambda_1^2\bar{k}_x^2+\lambda_2^2\bar{k}_y^2}, \notag\\
&\left.\frac{\partial \varphi(\mathbf{k} )}{ \partial k_y}\right|_{\mathbf{\bar{k}}}= -\frac{\lambda_1\lambda_2 \bar{k}_x}{ \lambda_1^2 \bar{k}_x^2+ \lambda_2^2\bar{k}_y^2}.
\label{AShift}
\end{align}
Here, $\mathbf{\bar{k}}=(\bar{k}_x, \bar{k}_y) $ is the momentum at which the atoms condense.  The lowest energy minima of the lower band depend on the anisotropy. When $\lambda_1< \lambda_2$, the two minima locate at $(\bar{k}_x, \bar{k}_y)=(0,\pm \lambda_2)$ [see Fig.~\ref{fig:anisotropy}(a1)]. They locate at $(\bar{k}_x, \bar{k}_y)=(\pm \lambda_1,0)$ when  $\lambda_1> \lambda_2$ [see Fig.~\ref{fig:anisotropy}(b1)].  With the miscible interactions, the BEC spontaneously chooses one of these two minima to condense. The ground state that spontaneously condenses at  $(\bar{k}_x, \bar{k}_y)=(0, -\lambda_2)$ for $\lambda_1< \lambda_2$ is demonstrated in Figs.~\ref{fig:anisotropy}(a2)--(a5). We obtain ground states by the imaginary-time evolution of the GP equation with $H'_\text{SOC}$.
From the single-particle prediction in Eq.~(\ref{AShift}), the phase separation of this ground state happens only along the $x$ direction, and the center of mass of the first component is $\lambda_1/(2\lambda_2^2)$ and that of the second component is $-\lambda_1/(2\lambda_2^2)$
[see the white stars in Figs.~\ref{fig:anisotropy}(a2) and \ref{fig:anisotropy}(a3)]. 
The density distributions shown in  Figs.~\ref{fig:anisotropy}(a2) and~\ref{fig:anisotropy}(a3) clearly indicate the phase separation following the predictions.  The ground state that spontaneously condenses at  $(\bar{k}_x, \bar{k}_y)=( -\lambda_1,0)$ for  $\lambda_1>\lambda_2$ is demonstrated in Figs.~\ref{fig:anisotropy}(b2)-(b5).  The single-particle mechanism in Eq.~(\ref{AShift}) predicts that, for this ground state, the separation happens along the $y$ direction and the centers of mass are $\mp \lambda_2/(2\lambda_1^2)$ for two components  [see the white stars in Figs.~\ref{fig:anisotropy}(b2) and \ref{fig:anisotropy}(b3)]. The results from the imaginary-time evolution shown in Figs.~\ref{fig:anisotropy}(b2) and~\ref{fig:anisotropy}(b3) match with the single-particle predictions.

\begin{figure}
\centering
\includegraphics[width=0.98\linewidth]{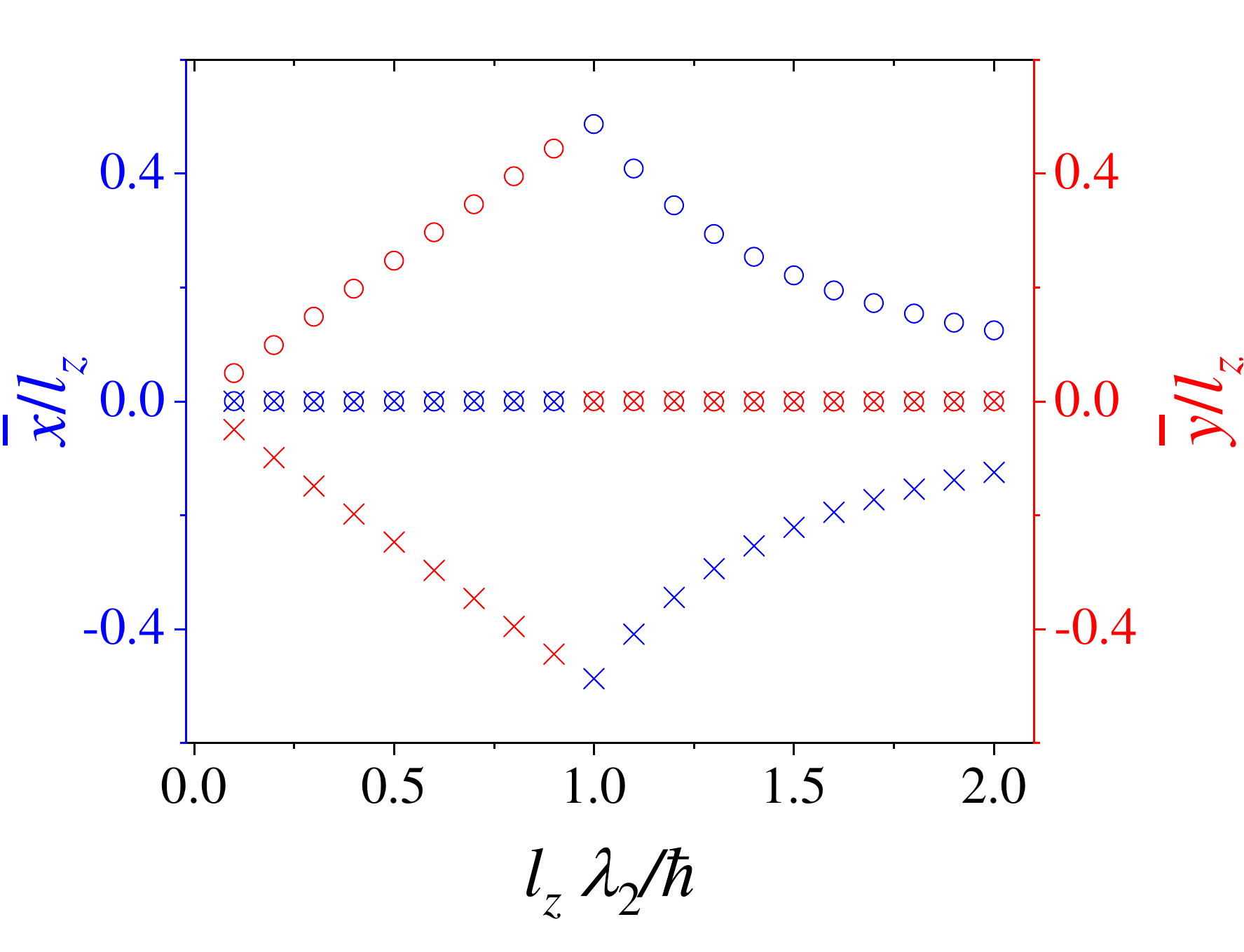}
\caption{Anisotropic-spin-orbit-coupling-induced phase separation as a function of $\lambda_2$ with a fixed $l_z\lambda_1/\hbar=1$. Circles are for the first component and crosses are for the second component. The blue (red) color represents separation along the $x$ ($y$) direction.
Other parameters are $\omega/\omega_z=0.1$, $g=12$ and $g_{12}=8$. }
\label{fig:anisotropy-com}
\end{figure}

These analyses showed that the center of mass of each component strongly depends on the ratio of the spin-orbit coupling strengths.
To reveal the dependence of  phase separation on $\lambda_2/\lambda_1$, 
we calculate ground states  with a fixed $\lambda_1$ and a changeable $\lambda_2$ by using the imaginary-time evolution. 
The results are summarized in Fig.~\ref{fig:anisotropy-com}, 
where the circles (crosses) represent the center of mass for the first (second) component.
For $\lambda_2<\lambda_1=1$, the phase separation occurs along the $y$ direction and $|\bar{y}|$ increases with the increase of $\lambda_2$ [see red circles and crosses in Fig.~\ref{fig:anisotropy-com}], 
while $\bar{x}$ is zero [see blue circles and crosses in Fig.~\ref{fig:anisotropy-com}]. When $\lambda_2=0$, the spin-orbit coupling becomes one-dimensional, there is no phase separation due to the absence of the relative phase.  
The results change for $\lambda_2>\lambda_1=1$ and the phase separation along the $x$ direction is observed.
In this case, the separation decreases with $\lambda_2$ increasing. For a very large $\lambda_2$, the separation disappears since the spin-orbit coupling effectively turns out to be one dimensional. The results in Fig.~\ref{fig:anisotropy-com} demonstrate that the maximum separation happens for $\lambda_1=\lambda_2$ which is Rashba spin-orbit coupling. This is also expected from the single-particle prediction in Eq.~(\ref{AShift}).

%%%%%%%%%%%%%%%%%%%%%%%%%%%%%%%%%%%%%%%%%%%%%
%%%%%%%%%%%%%%%%%%%%%%%%%%%%%%%%%%%%%%%%%%%%%
 
 %%%%%%%%%%%%%%%%%%%%%%%%%%%%%%%%%%%%%%%%%%%%%
 \begin{figure}[b]
 \centering
 \includegraphics[width=0.98\linewidth]{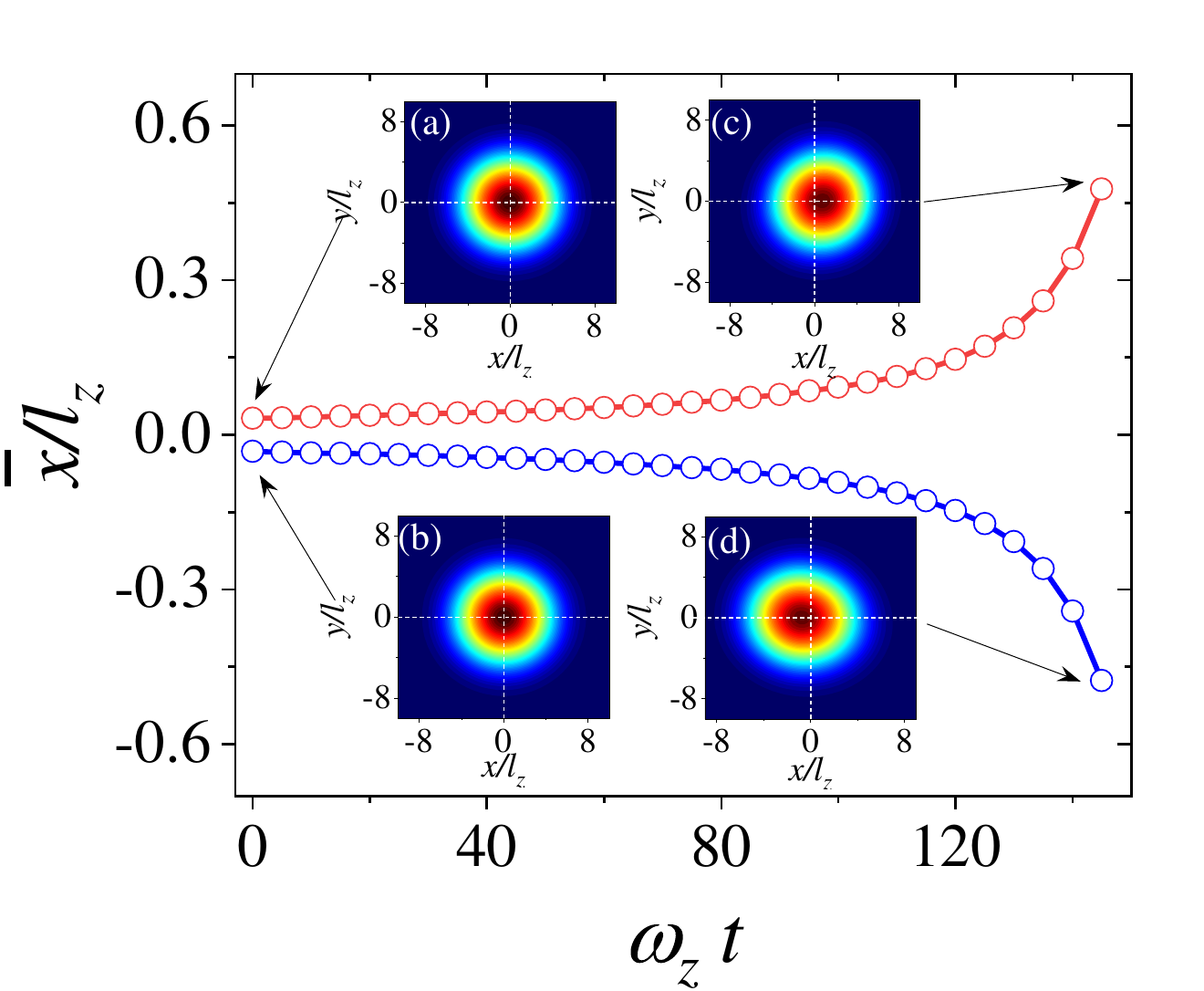}
 \caption{Adiabatic splitting dynamics of a trapped BEC with Rashba spin-orbit coupling by slowly switching off the linear coupling. The parameters are $\omega/\omega_z=0.1$, $g=12$, $g_{12}=8$, $\Omega_0/\omega_z=3$, and $\omega_z\tau_q=150$. The red (blue) dots represent the center of mass of the first (second) component.  Insets (a,b) [(c,d)] are density distributions of the first and second components at $t=0$ [$t=\tau_q$], respectively.}
 \label{fig:dynamics}
 \end{figure}
 %%%%%%%%%%%%%%%%%%%%%%%%%%%%%%%%%%%%%%%%%%%%%

\section{Adiabatic splitting dynamics}
\label{Adiabaticdynamics}

%%%%%%%%%%%%%%%%%%%%%%%%%%%%%%%%%%%%%%%%%%%%%
\begin{figure*}
\centering
\includegraphics[width=0.9\linewidth]{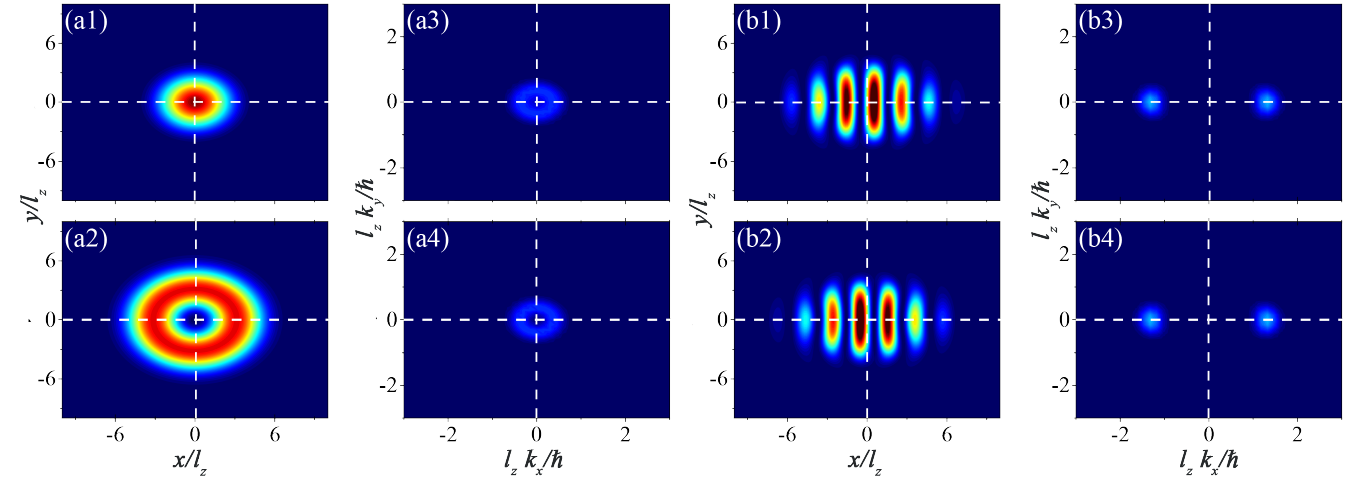}
\caption{Immiscible-interaction-induced phase-separated ground states in a trapped BEC with Rashba spin-orbit coupling. The parameters are $g=4$,  $g_{12}=8$ and $\omega/\omega_z=0.1$. (a1)--(a4) When $l_z\lambda/\hbar=0.5$ the ground state is a half-quantum vortex state. (a1) [a(3)] and (a2) [(a4)] are the coordinate (momentum) space density distributions of the first and second components respectively.  (b1)--(b4) When $l_z\lambda/\hbar=1.5$ the ground state is a stripe state. (b1) [b(3)] and (b2) [(b4)] are the coordinate [momentum] space density distributions of the first and second components respectively.}
\label{fig:strong-interspecies}
\end{figure*}
%%%%%%%%%%%%%%%%%%%%%%%%%%%%%%%%%%%%%%%%%%%%%
We showed the ground states of a trapped BEC with two-dimensional spin-orbit coupling and miscible interactions are phase-separated.  As an important application, we study the adiabatic dynamics of the phase separation. 
As pointed out by previous works, a linear coupling between the two components favors miscibility regardless of interactions~\cite{Gautam,Merhasin}.
Therefore, a miscible-to-immiscible transition may occur by decreasing the coupling. The adiabatic dynamics is stimulated by slowly switching off the linear coupling.  Theoretically, the process is described by the time-dependent GP equation, 
\begin{align}
i\frac{\partial\Psi}{\partial t}=\left[H_\text{SOC}+\Omega(t)\sigma_x+V+H_\text{int}\right] \Psi.
\label{eq:dynamics}
\end{align}
Here, $\Omega(t)\sigma_x $ represents the linear coupling between the two components, and can be experimentally achieved by using a radio-frequency coupling~\cite{Nicklas}.  The time-dependent Rabi frequency is 
\begin{equation}
\Omega(t)=\Omega_0(1-t/\tau_q),
\end{equation}
with $\Omega_0$ being the initial value of the linear coupling and $\tau_q$ is the quench duration.  At $t=0$, the presence of $\Omega_0$ greatly suppresses the ground-state phase separation. We obtain the ground state by the imaginary-time evolution of Eq.~(\ref{eq:dynamics}) with $\Omega(t)=\Omega_0$. A typical ground state is shown in the insets of Figs.~\ref{fig:dynamics} (a) and \ref{fig:dynamics}(b), and the separation between the two components is not obvious. Using this ground state as initial state, we evolve the time-dependent GP equation. The center-of-mass $\bar{x}$ for the two components is recorded during the time evolution in Fig.~\ref{fig:dynamics}.  By decreasing the linear coupling adiabatically, the separation between the two component gradually increases. When it is completely switched off, i.e., $t=\tau_q$, the separation is maximized [see the corresponding density distributions in insets (c) and (d)].  
The two components can realize a dynamically spatial splitting, which move along opposite directions.
Such adiabatic splitting dynamics are reminiscent of  a kind of  ``atomic spin Hall effect''~\cite{Beeler}.

\section{Immiscible interactions induced phase separation}
\label{Immiscibilty}

In all of the above, the interactions are miscible ($g>g_{12}$), which support atoms to condense at a particular momentum state. On the other hand, immiscible interactions ($g<g_{12}$) prefer a spatial separation between the two components to minimize the inter-component interactions proportional to $g_{12}$. In the presence of spin-orbit coupling, the immiscible-interaction-induced phase separation presents interesting features~\cite{Zhang2012,Hu2012,Ramachandhran}.  In Fig.~\ref{fig:strong-interspecies}, we show two different kinds of immiscible-interaction-induced phase separated ground states with different values of spin-orbit coupling strength $\lambda$. For $\lambda=0.5$, the ground state obtained by the imaginary-time evolution is a half-quantum vortex state which was first revealed in Refs.~\cite{Hu2012,Ramachandhran}. The first component distribution has a Gaussian shape [see Figs.~\ref{fig:strong-interspecies}(a1) and~\ref{fig:strong-interspecies}(a3)], and the second component is a vortex with a winding number $w=1$ [see Figs.~\ref{fig:strong-interspecies}(a2) and~\ref{fig:strong-interspecies}(a4)]. The first component is filled in the density dip of the second one, forming a spatial separation along the radial direction. For $\lambda=1.5$, the ground state becomes a stripe state, which was first revealed in Refs.~\cite{Wang, Ho2011}. The ground state condenses simultaneously at two different momenta [see Figs.~\ref{fig:strong-interspecies}(b3) and~\ref{fig:strong-interspecies}(b4)]. Such momentum occupation generates spatially periodic modulations in density distributions [Figs.~\ref{fig:strong-interspecies}(b1) and~\ref{fig:strong-interspecies}(b2)].  
Meanwhile, the stripes of the two components are spatially separated. 

We emphasize that phase separations induced by spin-orbit coupling and immiscible interaction have different physical origins. The spin-orbit-coupling-induced phase separation only works for a two-dimensional spin-orbit coupling.  
However, phase separations were also studied for a BEC with a one-dimensional spin-orbit coupling, the mechanism of which is different.  
In the pioneering spin-orbit-coupled experiment, the experimentalists observed a spatial separation between two dressed states with a Raman-induced spin-orbit coupling~\cite{Lin}. The spin-orbit coupling generated two energy minima, whose occupations can be considered as two dressed states.  In the dressed state space, the atomic interactions turned to be immiscible between the two dressed states in the presence of the Rabi frequency. The phase separation happened in the dressed state space due to immiscibility. 
In addition, the authors of Ref.~\cite{Gautam} revealed the existence of phase separation in a spin-1 BEC with the Raman-induced spin-orbit coupling. 
The single-particle Hamiltonian of the system is $H=(p_x+\lambda' F_z)^2/2 +\Omega' F_x +\epsilon F_z^2$. Here,  $F_{x,y,z}$ are the spin-1 Pauli matrices, $\lambda'$ is the spin-orbit coupling strength, $\Omega'$ is the Rabi frequency, and $\epsilon$ is the quadratic Zeeman shift. The spinor interactions include the density-density part with the coefficient $c_0$ and spin-spin part with the coefficient $c_2$. 
In particular, a very negative quadratic Zeeman shift $\epsilon=-\lambda^2/2$ was considered.
With such a large negative $\epsilon$, the occupation in the second component can be eliminated. The spinor only occupies the first and third components. Interestingly, the spinor interactions between the first and third components are immiscible for a negative spin-spin interaction ($c_2<0$). Different phase-separated states between the first and third component are due to immiscible interactions~\cite{Gautam} .

\section{Conclusion}
\label{conclusion}

In summary, we revealed the physical mechanism of spin-orbit-coupling-induced phase separation.  The mechanism, which is very different from the conventional immiscible-interaction-induced separation, is a complete single-particle effect of spin-orbit coupling. 
We analyzed the separation features in a trapped BEC with Rashba spin-orbit coupling and miscible interactions and studied the effects of the anisotropy of spin-orbit coupling on the separation.  All features can be explained by the single-particle mechanism. As an interesting application of the phase separation, we propose an adiabatic dynamics that can dynamically split two components spatially.

\section*{Acknowledgments}

This work was supported by National Natural Science Foundation of China with Grants No.12374247 and No. 11974235. 
H.L. acknowledges support from Okinawa  Institute of Science and Technology Graduate University.

\bibliography{RashbaPS}% Produces the bibliography via BibTeX.

%apsrev4-2.bst 2019-01-14 (MD) hand-edited version of apsrev4-1.bst
%Control: key (0)
%Control: author (8) initials jnrlst
%Control: editor formatted (1) identically to author
%Control: production of article title (0) allowed
%Control: page (0) single
%Control: year (1) truncated
%Control: production of eprint (0) enabled
\begin{thebibliography}{50}%
\makeatletter
\providecommand \@ifxundefined [1]{%
 \@ifx{#1\undefined}
}%
\providecommand \@ifnum [1]{%
 \ifnum #1\expandafter \@firstoftwo
 \else \expandafter \@secondoftwo
 \fi
}%
\providecommand \@ifx [1]{%
 \ifx #1\expandafter \@firstoftwo
 \else \expandafter \@secondoftwo
 \fi
}%
\providecommand \natexlab [1]{#1}%
\providecommand \enquote  [1]{``#1''}%
\providecommand \bibnamefont  [1]{#1}%
\providecommand \bibfnamefont [1]{#1}%
\providecommand \citenamefont [1]{#1}%
\providecommand \href@noop [0]{\@secondoftwo}%
\providecommand \href [0]{\begingroup \@sanitize@url \@href}%
\providecommand \@href[1]{\@@startlink{#1}\@@href}%
\providecommand \@@href[1]{\endgroup#1\@@endlink}%
\providecommand \@sanitize@url [0]{\catcode `\\12\catcode `\$12\catcode
  `\&12\catcode `\#12\catcode `\^12\catcode `\_12\catcode `\%12\relax}%
\providecommand \@@startlink[1]{}%
\providecommand \@@endlink[0]{}%
\providecommand \url  [0]{\begingroup\@sanitize@url \@url }%
\providecommand \@url [1]{\endgroup\@href {#1}{\urlprefix }}%
\providecommand \urlprefix  [0]{URL }%
\providecommand \Eprint [0]{\href }%
\providecommand \doibase [0]{https://doi.org/}%
\providecommand \selectlanguage [0]{\@gobble}%
\providecommand \bibinfo  [0]{\@secondoftwo}%
\providecommand \bibfield  [0]{\@secondoftwo}%
\providecommand \translation [1]{[#1]}%
\providecommand \BibitemOpen [0]{}%
\providecommand \bibitemStop [0]{}%
\providecommand \bibitemNoStop [0]{.\EOS\space}%
\providecommand \EOS [0]{\spacefactor3000\relax}%
\providecommand \BibitemShut  [1]{\csname bibitem#1\endcsname}%
\let\auto@bib@innerbib\@empty
%</preamble>
\bibitem [{\citenamefont {Sinova}\ \emph {et~al.}(2015)\citenamefont {Sinova},
  \citenamefont {Valenzuela}, \citenamefont {Wunderlich}, \citenamefont
  {Back},\ and\ \citenamefont {Jungwirth}}]{Sinova}%
  \BibitemOpen
  \bibfield  {author} {\bibinfo {author} {\bibfnamefont {J.}~\bibnamefont
  {Sinova}}, \bibinfo {author} {\bibfnamefont {S.~O.}\ \bibnamefont
  {Valenzuela}}, \bibinfo {author} {\bibfnamefont {J.}~\bibnamefont
  {Wunderlich}}, \bibinfo {author} {\bibfnamefont {C.~H.}\ \bibnamefont
  {Back}},\ and\ \bibinfo {author} {\bibfnamefont {T.}~\bibnamefont
  {Jungwirth}},\ }\bibfield  {title} {\bibinfo {title} {Spin {H}all effects},\
  }\href {https://doi.org/10.1103/RevModPhys.87.1213} {\bibfield  {journal}
  {\bibinfo  {journal} {Rev. Mod. Phys.}\ }\textbf {\bibinfo {volume} {87}},\
  \bibinfo {pages} {1213} (\bibinfo {year} {2015})}\BibitemShut {NoStop}%
\bibitem [{\citenamefont {Shin}\ \emph {et~al.}(2006)\citenamefont {Shin},
  \citenamefont {Zwierlein}, \citenamefont {Schunck}, \citenamefont
  {Schirotzek},\ and\ \citenamefont {Ketterle}}]{Shin}%
  \BibitemOpen
  \bibfield  {author} {\bibinfo {author} {\bibfnamefont {Y.}~\bibnamefont
  {Shin}}, \bibinfo {author} {\bibfnamefont {M.~W.}\ \bibnamefont {Zwierlein}},
  \bibinfo {author} {\bibfnamefont {C.~H.}\ \bibnamefont {Schunck}}, \bibinfo
  {author} {\bibfnamefont {A.}~\bibnamefont {Schirotzek}},\ and\ \bibinfo
  {author} {\bibfnamefont {W.}~\bibnamefont {Ketterle}},\ }\bibfield  {title}
  {\bibinfo {title} {Observation of {P}hase {S}eparation in a {S}trongly
  {I}nteracting {I}mbalanced {F}ermi {G}as},\ }\href
  {https://doi.org/10.1103/PhysRevLett.97.030401} {\bibfield  {journal}
  {\bibinfo  {journal} {Phys. Rev. Lett.}\ }\textbf {\bibinfo {volume} {97}},\
  \bibinfo {pages} {030401} (\bibinfo {year} {2006})}\BibitemShut {NoStop}%
\bibitem [{\citenamefont {Papp}\ \emph {et~al.}(2008)\citenamefont {Papp},
  \citenamefont {Pino},\ and\ \citenamefont {Wieman}}]{Papp}%
  \BibitemOpen
  \bibfield  {author} {\bibinfo {author} {\bibfnamefont {S.~B.}\ \bibnamefont
  {Papp}}, \bibinfo {author} {\bibfnamefont {J.~M.}\ \bibnamefont {Pino}},\
  and\ \bibinfo {author} {\bibfnamefont {C.~E.}\ \bibnamefont {Wieman}},\
  }\bibfield  {title} {\bibinfo {title} {Tunable {M}iscibility in a
  {Dual}-{Species} {Bose}-{Einstein} {Condensate}},\ }\href
  {https://doi.org/10.1103/PhysRevLett.101.040402} {\bibfield  {journal}
  {\bibinfo  {journal} {Phys. Rev. Lett.}\ }\textbf {\bibinfo {volume} {101}},\
  \bibinfo {pages} {040402} (\bibinfo {year} {2008})}\BibitemShut {NoStop}%
\bibitem [{\citenamefont {Thalhammer}\ \emph {et~al.}(2008)\citenamefont
  {Thalhammer}, \citenamefont {Barontini}, \citenamefont {De~Sarlo},
  \citenamefont {Catani}, \citenamefont {Minardi},\ and\ \citenamefont
  {Inguscio}}]{Thalhammer}%
  \BibitemOpen
  \bibfield  {author} {\bibinfo {author} {\bibfnamefont {G.}~\bibnamefont
  {Thalhammer}}, \bibinfo {author} {\bibfnamefont {G.}~\bibnamefont
  {Barontini}}, \bibinfo {author} {\bibfnamefont {L.}~\bibnamefont {De~Sarlo}},
  \bibinfo {author} {\bibfnamefont {J.}~\bibnamefont {Catani}}, \bibinfo
  {author} {\bibfnamefont {F.}~\bibnamefont {Minardi}},\ and\ \bibinfo {author}
  {\bibfnamefont {M.}~\bibnamefont {Inguscio}},\ }\bibfield  {title} {\bibinfo
  {title} {{D}ouble {S}pecies {B}ose-{E}instein {C}ondensate with {T}unable
  {I}nterspecies {I}nteractions},\ }\href
  {https://doi.org/10.1103/PhysRevLett.100.210402} {\bibfield  {journal}
  {\bibinfo  {journal} {Phys. Rev. Lett.}\ }\textbf {\bibinfo {volume} {100}},\
  \bibinfo {pages} {210402} (\bibinfo {year} {2008})}\BibitemShut {NoStop}%
\bibitem [{\citenamefont {Tojo}\ \emph {et~al.}(2010)\citenamefont {Tojo},
  \citenamefont {Taguchi}, \citenamefont {Masuyama}, \citenamefont {Hayashi},
  \citenamefont {Saito},\ and\ \citenamefont {Hirano}}]{Tojo}%
  \BibitemOpen
  \bibfield  {author} {\bibinfo {author} {\bibfnamefont {S.}~\bibnamefont
  {Tojo}}, \bibinfo {author} {\bibfnamefont {Y.}~\bibnamefont {Taguchi}},
  \bibinfo {author} {\bibfnamefont {Y.}~\bibnamefont {Masuyama}}, \bibinfo
  {author} {\bibfnamefont {T.}~\bibnamefont {Hayashi}}, \bibinfo {author}
  {\bibfnamefont {H.}~\bibnamefont {Saito}},\ and\ \bibinfo {author}
  {\bibfnamefont {T.}~\bibnamefont {Hirano}},\ }\bibfield  {title} {\bibinfo
  {title} {Controlling phase separation of binary {B}ose-{E}instein condensates
  via mixed-spin-channel {F}eshbach resonance},\ }\href
  {https://doi.org/10.1103/PhysRevA.82.033609} {\bibfield  {journal} {\bibinfo
  {journal} {Phys. Rev. A}\ }\textbf {\bibinfo {volume} {82}},\ \bibinfo
  {pages} {033609} (\bibinfo {year} {2010})}\BibitemShut {NoStop}%
\bibitem [{\citenamefont {Nicklas}\ \emph {et~al.}(2011)\citenamefont
  {Nicklas}, \citenamefont {Strobel}, \citenamefont {Zibold}, \citenamefont
  {Gross}, \citenamefont {Malomed}, \citenamefont {Kevrekidis},\ and\
  \citenamefont {Oberthaler}}]{Nicklas}%
  \BibitemOpen
  \bibfield  {author} {\bibinfo {author} {\bibfnamefont {E.}~\bibnamefont
  {Nicklas}}, \bibinfo {author} {\bibfnamefont {H.}~\bibnamefont {Strobel}},
  \bibinfo {author} {\bibfnamefont {T.}~\bibnamefont {Zibold}}, \bibinfo
  {author} {\bibfnamefont {C.}~\bibnamefont {Gross}}, \bibinfo {author}
  {\bibfnamefont {B.~A.}\ \bibnamefont {Malomed}}, \bibinfo {author}
  {\bibfnamefont {P.~G.}\ \bibnamefont {Kevrekidis}},\ and\ \bibinfo {author}
  {\bibfnamefont {M.~K.}\ \bibnamefont {Oberthaler}},\ }\bibfield  {title}
  {\bibinfo {title} {Rabi {F}lopping {I}nduces {S}patial {D}emixing
  {D}ynamics},\ }\href {https://doi.org/10.1103/PhysRevLett.107.193001}
  {\bibfield  {journal} {\bibinfo  {journal} {Phys. Rev. Lett.}\ }\textbf
  {\bibinfo {volume} {107}},\ \bibinfo {pages} {193001} (\bibinfo {year}
  {2011})}\BibitemShut {NoStop}%
\bibitem [{\citenamefont {Jim\'enez-Garc\'ia}\ \emph
  {et~al.}(2019)\citenamefont {Jim\'enez-Garc\'ia}, \citenamefont {Invernizzi},
  \citenamefont {Evrard}, \citenamefont {Frapolli}, \citenamefont {Dalibard},\
  and\ \citenamefont {Gerbier}}]{Ota}%
  \BibitemOpen
  \bibfield  {author} {\bibinfo {author} {\bibfnamefont {K.}~\bibnamefont
  {Jim\'enez-Garc\'ia}}, \bibinfo {author} {\bibfnamefont {A.}~\bibnamefont
  {Invernizzi}}, \bibinfo {author} {\bibfnamefont {B.}~\bibnamefont {Evrard}},
  \bibinfo {author} {\bibfnamefont {C.}~\bibnamefont {Frapolli}}, \bibinfo
  {author} {\bibfnamefont {J.}~\bibnamefont {Dalibard}},\ and\ \bibinfo
  {author} {\bibfnamefont {F.}~\bibnamefont {Gerbier}},\ }\bibfield  {title}
  {\bibinfo {title} {Spontaneous formation and relaxation of spin domains in
  antiferromagnetic spin-1 condensates},\ }\href
  {https://doi.org/10.1038/s41467-019-08505-6} {\bibfield  {journal} {\bibinfo
  {journal} {Nat. Commun.}\ }\textbf {\bibinfo {volume} {10}},\ \bibinfo
  {pages} {1422} (\bibinfo {year} {2019})}\BibitemShut {NoStop}%
\bibitem [{\citenamefont {He}\ \emph {et~al.}(2020)\citenamefont {He},
  \citenamefont {Gao},\ and\ \citenamefont {Yu}}]{HeL}%
  \BibitemOpen
  \bibfield  {author} {\bibinfo {author} {\bibfnamefont {L.}~\bibnamefont
  {He}}, \bibinfo {author} {\bibfnamefont {P.}~\bibnamefont {Gao}},\ and\
  \bibinfo {author} {\bibfnamefont {Z.-Q.}\ \bibnamefont {Yu}},\ }\bibfield
  {title} {\bibinfo {title} {Normal-{S}uperfluid {P}hase {S}eparation in
  {S}pin-{H}alf {B}osons at {F}inite {T}emperature},\ }\href
  {https://doi.org/10.1103/PhysRevLett.125.055301} {\bibfield  {journal}
  {\bibinfo  {journal} {Phys. Rev. Lett.}\ }\textbf {\bibinfo {volume} {125}},\
  \bibinfo {pages} {055301} (\bibinfo {year} {2020})}\BibitemShut {NoStop}%
\bibitem [{\citenamefont {Trippenbach}\ \emph {et~al.}(2000)\citenamefont
  {Trippenbach}, \citenamefont {G\'oral}, \citenamefont {Rzazewski},
  \citenamefont {Malomed},\ and\ \citenamefont {Band}}]{Trippenbach}%
  \BibitemOpen
  \bibfield  {author} {\bibinfo {author} {\bibfnamefont {M.}~\bibnamefont
  {Trippenbach}}, \bibinfo {author} {\bibfnamefont {K.}~\bibnamefont
  {G\'oral}}, \bibinfo {author} {\bibfnamefont {K.}~\bibnamefont {Rzazewski}},
  \bibinfo {author} {\bibfnamefont {B.}~\bibnamefont {Malomed}},\ and\ \bibinfo
  {author} {\bibfnamefont {Y.~B.}\ \bibnamefont {Band}},\ }\bibfield  {title}
  {\bibinfo {title} {Structure of binary {B}ose-{E}instein condensates},\
  }\href {https://doi.org/10.1088/0953-4075/33/19/314} {\bibfield  {journal}
  {\bibinfo  {journal} {J. Phys. B}\ }\textbf {\bibinfo {volume} {33}},\
  \bibinfo {pages} {4017} (\bibinfo {year} {2000})}\BibitemShut {NoStop}%
\bibitem [{\citenamefont {Petrov}(2015)}]{Petrov2015}%
  \BibitemOpen
  \bibfield  {author} {\bibinfo {author} {\bibfnamefont {D.~S.}\ \bibnamefont
  {Petrov}},\ }\bibfield  {title} {\bibinfo {title} {Quantum {M}echanical
  {S}tabilization of a {C}ollapsing {B}ose-{B}ose {M}ixture},\ }\href
  {https://doi.org/10.1103/PhysRevLett.115.155302} {\bibfield  {journal}
  {\bibinfo  {journal} {Phys. Rev. Lett.}\ }\textbf {\bibinfo {volume} {115}},\
  \bibinfo {pages} {155302} (\bibinfo {year} {2015})}\BibitemShut {NoStop}%
\bibitem [{\citenamefont {Shrestha}\ \emph {et~al.}(2009)\citenamefont
  {Shrestha}, \citenamefont {Javanainen},\ and\ \citenamefont
  {Ruostekoski}}]{Shrestha}%
  \BibitemOpen
  \bibfield  {author} {\bibinfo {author} {\bibfnamefont {U.}~\bibnamefont
  {Shrestha}}, \bibinfo {author} {\bibfnamefont {J.}~\bibnamefont
  {Javanainen}},\ and\ \bibinfo {author} {\bibfnamefont {J.}~\bibnamefont
  {Ruostekoski}},\ }\bibfield  {title} {\bibinfo {title} {Pulsating and
  {P}ersistent {V}ector {S}olitons in a {B}ose-{E}instein {C}ondensate in a
  {L}attice upon {P}hase {S}eparation {I}nstability},\ }\href
  {https://doi.org/10.1103/PhysRevLett.103.190401} {\bibfield  {journal}
  {\bibinfo  {journal} {Phys. Rev. Lett.}\ }\textbf {\bibinfo {volume} {103}},\
  \bibinfo {pages} {190401} (\bibinfo {year} {2009})}\BibitemShut {NoStop}%
\bibitem [{\citenamefont {Law}\ \emph {et~al.}(2010)\citenamefont {Law},
  \citenamefont {Kevrekidis},\ and\ \citenamefont {Tuckerman}}]{Law}%
  \BibitemOpen
  \bibfield  {author} {\bibinfo {author} {\bibfnamefont {K.~J.~H.}\
  \bibnamefont {Law}}, \bibinfo {author} {\bibfnamefont {P.~G.}\ \bibnamefont
  {Kevrekidis}},\ and\ \bibinfo {author} {\bibfnamefont {L.~S.}\ \bibnamefont
  {Tuckerman}},\ }\bibfield  {title} {\bibinfo {title} {Stable
  {V}ortex--{B}right--{S}oliton structures in {T}wo-{C}omponent
  {B}ose-{E}instein {C}ondensates},\ }\href
  {https://doi.org/10.1103/PhysRevLett.105.160405} {\bibfield  {journal}
  {\bibinfo  {journal} {Phys. Rev. Lett.}\ }\textbf {\bibinfo {volume} {105}},\
  \bibinfo {pages} {160405} (\bibinfo {year} {2010})}\BibitemShut {NoStop}%
\bibitem [{\citenamefont {Lee}\ \emph {et~al.}(2016)\citenamefont {Lee},
  \citenamefont {J\o{}rgensen}, \citenamefont {Liu}, \citenamefont {Wacker},
  \citenamefont {Arlt},\ and\ \citenamefont {Proukakis}}]{LeeKL}%
  \BibitemOpen
  \bibfield  {author} {\bibinfo {author} {\bibfnamefont {K.~L.}\ \bibnamefont
  {Lee}}, \bibinfo {author} {\bibfnamefont {N.~B.}\ \bibnamefont
  {J\o{}rgensen}}, \bibinfo {author} {\bibfnamefont {I.-K.}\ \bibnamefont
  {Liu}}, \bibinfo {author} {\bibfnamefont {L.}~\bibnamefont {Wacker}},
  \bibinfo {author} {\bibfnamefont {J.~J.}\ \bibnamefont {Arlt}},\ and\
  \bibinfo {author} {\bibfnamefont {N.~P.}\ \bibnamefont {Proukakis}},\
  }\bibfield  {title} {\bibinfo {title} {Phase separation and dynamics of
  two-component {Bose}-{Einstein} condensates},\ }\href
  {https://doi.org/10.1103/PhysRevA.94.013602} {\bibfield  {journal} {\bibinfo
  {journal} {Phys. Rev. A}\ }\textbf {\bibinfo {volume} {94}},\ \bibinfo
  {pages} {013602} (\bibinfo {year} {2016})}\BibitemShut {NoStop}%
\bibitem [{\citenamefont {Roy}\ \emph {et~al.}(2021)\citenamefont {Roy},
  \citenamefont {Ota}, \citenamefont {Recati},\ and\ \citenamefont
  {Dalfovo}}]{Roy}%
  \BibitemOpen
  \bibfield  {author} {\bibinfo {author} {\bibfnamefont {A.}~\bibnamefont
  {Roy}}, \bibinfo {author} {\bibfnamefont {M.}~\bibnamefont {Ota}}, \bibinfo
  {author} {\bibfnamefont {A.}~\bibnamefont {Recati}},\ and\ \bibinfo {author}
  {\bibfnamefont {F.}~\bibnamefont {Dalfovo}},\ }\bibfield  {title} {\bibinfo
  {title} {Finite-temperature spin dynamics of a two-dimensional {Bose}-{Bose}
  atomic mixture},\ }\href {https://doi.org/10.1103/PhysRevResearch.3.013161}
  {\bibfield  {journal} {\bibinfo  {journal} {Phys. Rev. Res.}\ }\textbf
  {\bibinfo {volume} {3}},\ \bibinfo {pages} {013161} (\bibinfo {year}
  {2021})}\BibitemShut {NoStop}%
\bibitem [{\citenamefont {Eto}\ \emph {et~al.}(2016)\citenamefont {Eto},
  \citenamefont {Takahashi}, \citenamefont {Kunimi}, \citenamefont {Saito},\
  and\ \citenamefont {Hirano}}]{Eto}%
  \BibitemOpen
  \bibfield  {author} {\bibinfo {author} {\bibfnamefont {Y.}~\bibnamefont
  {Eto}}, \bibinfo {author} {\bibfnamefont {M.}~\bibnamefont {Takahashi}},
  \bibinfo {author} {\bibfnamefont {M.}~\bibnamefont {Kunimi}}, \bibinfo
  {author} {\bibfnamefont {H.}~\bibnamefont {Saito}},\ and\ \bibinfo {author}
  {\bibfnamefont {T.}~\bibnamefont {Hirano}},\ }\bibfield  {title} {\bibinfo
  {title} {Nonequilibrium dynamics induced by miscible–immiscible transition
  in binary {B}ose–{E}instein condensates},\ }\href
  {https://doi.org/10.1088/1367-2630/18/7/073029} {\bibfield  {journal}
  {\bibinfo  {journal} {New J. Phys.}\ }\textbf {\bibinfo {volume} {18}},\
  \bibinfo {pages} {073029} (\bibinfo {year} {2016})}\BibitemShut {NoStop}%
\bibitem [{\citenamefont {Bernier}\ \emph {et~al.}(2014)\citenamefont
  {Bernier}, \citenamefont {Dalla~Torre},\ and\ \citenamefont
  {Demler}}]{Bernier}%
  \BibitemOpen
  \bibfield  {author} {\bibinfo {author} {\bibfnamefont {N.~R.}\ \bibnamefont
  {Bernier}}, \bibinfo {author} {\bibfnamefont {E.~G.}\ \bibnamefont
  {Dalla~Torre}},\ and\ \bibinfo {author} {\bibfnamefont {E.}~\bibnamefont
  {Demler}},\ }\bibfield  {title} {\bibinfo {title} {Unstable {Avoided}
  {Crossing} in {Coupled} {Spinor} {Condensates}},\ }\href
  {https://doi.org/10.1103/PhysRevLett.113.065303} {\bibfield  {journal}
  {\bibinfo  {journal} {Phys. Rev. Lett.}\ }\textbf {\bibinfo {volume} {113}},\
  \bibinfo {pages} {065303} (\bibinfo {year} {2014})}\BibitemShut {NoStop}%
\bibitem [{\citenamefont {Shirley}\ \emph {et~al.}(2014)\citenamefont
  {Shirley}, \citenamefont {Anderson}, \citenamefont {Clark},\ and\
  \citenamefont {Wilson}}]{Shirley}%
  \BibitemOpen
  \bibfield  {author} {\bibinfo {author} {\bibfnamefont {W.~E.}\ \bibnamefont
  {Shirley}}, \bibinfo {author} {\bibfnamefont {B.~M.}\ \bibnamefont
  {Anderson}}, \bibinfo {author} {\bibfnamefont {C.~W.}\ \bibnamefont
  {Clark}},\ and\ \bibinfo {author} {\bibfnamefont {R.~M.}\ \bibnamefont
  {Wilson}},\ }\bibfield  {title} {\bibinfo {title} {Half-{Q}uantum {V}ortex
  {M}olecules in a {B}inary {D}ipolar {B}ose {G}as},\ }\href
  {https://doi.org/10.1103/PhysRevLett.113.165301} {\bibfield  {journal}
  {\bibinfo  {journal} {Phys. Rev. Lett.}\ }\textbf {\bibinfo {volume} {113}},\
  \bibinfo {pages} {165301} (\bibinfo {year} {2014})}\BibitemShut {NoStop}%
\bibitem [{\citenamefont {Mistakidis}\ \emph {et~al.}(2018)\citenamefont
  {Mistakidis}, \citenamefont {Katsimiga}, \citenamefont {Kevrekidis},\ and\
  \citenamefont {Schmelcher}}]{Mistakidis}%
  \BibitemOpen
  \bibfield  {author} {\bibinfo {author} {\bibfnamefont {S.~I.}\ \bibnamefont
  {Mistakidis}}, \bibinfo {author} {\bibfnamefont {G.~C.}\ \bibnamefont
  {Katsimiga}}, \bibinfo {author} {\bibfnamefont {P.~G.}\ \bibnamefont
  {Kevrekidis}},\ and\ \bibinfo {author} {\bibfnamefont {P.}~\bibnamefont
  {Schmelcher}},\ }\bibfield  {title} {\bibinfo {title} {Correlation effects in
  the quench-induced phase separation dynamics of a two species ultracold
  quantum gas},\ }\href {https://doi.org/10.1088/1367-2630/aabc6a} {\bibfield
  {journal} {\bibinfo  {journal} {New J. Phys.}\ }\textbf {\bibinfo {volume}
  {20}},\ \bibinfo {pages} {043052} (\bibinfo {year} {2018})}\BibitemShut
  {NoStop}%
\bibitem [{\citenamefont {Lous}\ \emph {et~al.}(2018)\citenamefont {Lous},
  \citenamefont {Fritsche}, \citenamefont {Jag}, \citenamefont {Lehmann},
  \citenamefont {Kirilov}, \citenamefont {Huang},\ and\ \citenamefont
  {Grimm}}]{Lous}%
  \BibitemOpen
  \bibfield  {author} {\bibinfo {author} {\bibfnamefont {R.~S.}\ \bibnamefont
  {Lous}}, \bibinfo {author} {\bibfnamefont {I.}~\bibnamefont {Fritsche}},
  \bibinfo {author} {\bibfnamefont {M.}~\bibnamefont {Jag}}, \bibinfo {author}
  {\bibfnamefont {F.}~\bibnamefont {Lehmann}}, \bibinfo {author} {\bibfnamefont
  {E.}~\bibnamefont {Kirilov}}, \bibinfo {author} {\bibfnamefont
  {B.}~\bibnamefont {Huang}},\ and\ \bibinfo {author} {\bibfnamefont
  {R.}~\bibnamefont {Grimm}},\ }\bibfield  {title} {\bibinfo {title} {Probing
  the {I}nterface of a {P}hase-{S}eparated {S}tate in a {R}epulsive
  {B}ose-{F}ermi {M}ixture},\ }\href
  {https://doi.org/10.1103/PhysRevLett.120.243403} {\bibfield  {journal}
  {\bibinfo  {journal} {Phys. Rev. Lett.}\ }\textbf {\bibinfo {volume} {120}},\
  \bibinfo {pages} {243403} (\bibinfo {year} {2018})}\BibitemShut {NoStop}%
\bibitem [{\citenamefont {Lin}\ \emph {et~al.}(2011)\citenamefont {Lin},
  \citenamefont {Jiménez-García},\ and\ \citenamefont {Spielman}}]{Lin}%
  \BibitemOpen
  \bibfield  {author} {\bibinfo {author} {\bibfnamefont {Y.-J.}\ \bibnamefont
  {Lin}}, \bibinfo {author} {\bibfnamefont {K.}~\bibnamefont
  {Jiménez-García}},\ and\ \bibinfo {author} {\bibfnamefont {I.~B.}\
  \bibnamefont {Spielman}},\ }\bibfield  {title} {\bibinfo {title}
  {Spin--orbit-coupled {Bose}–{Einstein} condensates},\ }\href
  {https://doi.org/10.1038/nature09887} {\bibfield  {journal} {\bibinfo
  {journal} {Nature (London)}\ }\textbf {\bibinfo {volume} {471}},\ \bibinfo
  {pages} {83} (\bibinfo {year} {2011})}\BibitemShut {NoStop}%
\bibitem [{\citenamefont {Goldman}\ \emph {et~al.}(2014)\citenamefont
  {Goldman}, \citenamefont {Juzeli{\={u}}nas}, \citenamefont {\"Ohberg},\ and\
  \citenamefont {Spielman}}]{Goldman}%
  \BibitemOpen
  \bibfield  {author} {\bibinfo {author} {\bibfnamefont {N.}~\bibnamefont
  {Goldman}}, \bibinfo {author} {\bibfnamefont {G.}~\bibnamefont
  {Juzeli{\={u}}nas}}, \bibinfo {author} {\bibfnamefont {P.}~\bibnamefont
  {\"Ohberg}},\ and\ \bibinfo {author} {\bibfnamefont {I.~B.}\ \bibnamefont
  {Spielman}},\ }\bibfield  {title} {\bibinfo {title} {Light-induced gauge
  fields for ultracold atoms},\ }\href
  {https://doi.org/10.1088/0034-4885/77/12/126401} {\bibfield  {journal}
  {\bibinfo  {journal} {Rep. Prog. Phys.}\ }\textbf {\bibinfo {volume} {77}},\
  \bibinfo {pages} {126401} (\bibinfo {year} {2014})}\BibitemShut {NoStop}%
\bibitem [{\citenamefont {Zhai}(2015)}]{Zhai2015}%
  \BibitemOpen
  \bibfield  {author} {\bibinfo {author} {\bibfnamefont {H.}~\bibnamefont
  {Zhai}},\ }\bibfield  {title} {\bibinfo {title} {Degenerate quantum gases
  with spin{\textendash}orbit coupling: a review},\ }\href
  {https://doi.org/10.1088/0034-4885/78/2/026001} {\bibfield  {journal}
  {\bibinfo  {journal} {Rep. Prog. Phys.}\ }\textbf {\bibinfo {volume} {78}},\
  \bibinfo {pages} {026001} (\bibinfo {year} {2015})}\BibitemShut {NoStop}%
\bibitem [{\citenamefont {Zhang}\ \emph {et~al.}(2016)\citenamefont {Zhang},
  \citenamefont {Mossman}, \citenamefont {Busch}, \citenamefont {Engels},\ and\
  \citenamefont {Zhang}}]{Zhang2016}%
  \BibitemOpen
  \bibfield  {author} {\bibinfo {author} {\bibfnamefont {Y.}~\bibnamefont
  {Zhang}}, \bibinfo {author} {\bibfnamefont {M.~E.}\ \bibnamefont {Mossman}},
  \bibinfo {author} {\bibfnamefont {T.}~\bibnamefont {Busch}}, \bibinfo
  {author} {\bibfnamefont {P.}~\bibnamefont {Engels}},\ and\ \bibinfo {author}
  {\bibfnamefont {C.}~\bibnamefont {Zhang}},\ }\bibfield  {title} {\bibinfo
  {title} {Properties of spin–orbit-coupled {B}ose–{E}instein
  condensates},\ }\href {https://doi.org/10.1007/s11467-016-0560-y} {\bibfield
  {journal} {\bibinfo  {journal} {Front. Phys.}\ }\textbf {\bibinfo {volume}
  {11}},\ \bibinfo {pages} {118103} (\bibinfo {year} {2016})}\BibitemShut
  {NoStop}%
\bibitem [{\citenamefont {Wu}\ \emph {et~al.}(2016)\citenamefont {Wu},
  \citenamefont {Zhang}, \citenamefont {Sun}, \citenamefont {Xu}, \citenamefont
  {Wang}, \citenamefont {Ji}, \citenamefont {Deng}, \citenamefont {Chen},
  \citenamefont {Liu},\ and\ \citenamefont {Pan}}]{WuZ}%
  \BibitemOpen
  \bibfield  {author} {\bibinfo {author} {\bibfnamefont {Z.}~\bibnamefont
  {Wu}}, \bibinfo {author} {\bibfnamefont {L.}~\bibnamefont {Zhang}}, \bibinfo
  {author} {\bibfnamefont {W.}~\bibnamefont {Sun}}, \bibinfo {author}
  {\bibfnamefont {X.-T.}\ \bibnamefont {Xu}}, \bibinfo {author} {\bibfnamefont
  {B.-Z.}\ \bibnamefont {Wang}}, \bibinfo {author} {\bibfnamefont {S.-C.}\
  \bibnamefont {Ji}}, \bibinfo {author} {\bibfnamefont {Y.}~\bibnamefont
  {Deng}}, \bibinfo {author} {\bibfnamefont {S.}~\bibnamefont {Chen}}, \bibinfo
  {author} {\bibfnamefont {X.-J.}\ \bibnamefont {Liu}},\ and\ \bibinfo {author}
  {\bibfnamefont {J.-W.}\ \bibnamefont {Pan}},\ }\bibfield  {title} {\bibinfo
  {title} {Realization of two-dimensional spin-orbit coupling for
  {Bose}-{Einstein} condensates},\ }\href
  {https://doi.org/10.1126/science.aaf6689} {\bibfield  {journal} {\bibinfo
  {journal} {Science}\ }\textbf {\bibinfo {volume} {354}},\ \bibinfo {pages}
  {83} (\bibinfo {year} {2016})}\BibitemShut {NoStop}%
\bibitem [{\citenamefont {Vald\'es-Curiel}\ \emph {et~al.}(2021)\citenamefont
  {Vald\'es-Curiel}, \citenamefont {Trypogeorgos}, \citenamefont {Liang},
  \citenamefont {Anderson},\ and\ \citenamefont {Spielman}}]{Valdes}%
  \BibitemOpen
  \bibfield  {author} {\bibinfo {author} {\bibfnamefont {A.}~\bibnamefont
  {Vald\'es-Curiel}}, \bibinfo {author} {\bibfnamefont {D.}~\bibnamefont
  {Trypogeorgos}}, \bibinfo {author} {\bibfnamefont {Q.-Y.}\ \bibnamefont
  {Liang}}, \bibinfo {author} {\bibfnamefont {R.~P.}\ \bibnamefont
  {Anderson}},\ and\ \bibinfo {author} {\bibfnamefont {I.~B.}\ \bibnamefont
  {Spielman}},\ }\bibfield  {title} {\bibinfo {title} {Topological features
  without a lattice in {Rashba} spin-orbit coupled atoms},\ }\href
  {https://doi.org/10.1038/s41467-020-20762-4} {\bibfield  {journal} {\bibinfo
  {journal} {Nat. Commun.}\ }\textbf {\bibinfo {volume} {12}},\ \bibinfo
  {pages} {593} (\bibinfo {year} {2021})}\BibitemShut {NoStop}%
\bibitem [{\citenamefont {Li}\ \emph {et~al.}(2012)\citenamefont {Li},
  \citenamefont {Pitaevskii},\ and\ \citenamefont {Stringari}}]{LiY}%
  \BibitemOpen
  \bibfield  {author} {\bibinfo {author} {\bibfnamefont {Y.}~\bibnamefont
  {Li}}, \bibinfo {author} {\bibfnamefont {L.~P.}\ \bibnamefont {Pitaevskii}},\
  and\ \bibinfo {author} {\bibfnamefont {S.}~\bibnamefont {Stringari}},\
  }\bibfield  {title} {\bibinfo {title} {Quantum {Tricriticality} and {Phase}
  {Transitions} in {Spin}-{Orbit} {Coupled} {Bose}-{Einstein} {Condensates}},\
  }\href {https://doi.org/10.1103/PhysRevLett.108.225301} {\bibfield  {journal}
  {\bibinfo  {journal} {Phys. Rev. Lett.}\ }\textbf {\bibinfo {volume} {108}},\
  \bibinfo {pages} {225301} (\bibinfo {year} {2012})}\BibitemShut {NoStop}%
\bibitem [{\citenamefont {Manchon}\ \emph {et~al.}(2015)\citenamefont
  {Manchon}, \citenamefont {Koo}, \citenamefont {Nitta}, \citenamefont
  {Frolov},\ and\ \citenamefont {Duine}}]{Manchon}%
  \BibitemOpen
  \bibfield  {author} {\bibinfo {author} {\bibfnamefont {A.}~\bibnamefont
  {Manchon}}, \bibinfo {author} {\bibfnamefont {H.~C.}\ \bibnamefont {Koo}},
  \bibinfo {author} {\bibfnamefont {J.}~\bibnamefont {Nitta}}, \bibinfo
  {author} {\bibfnamefont {S.~M.}\ \bibnamefont {Frolov}},\ and\ \bibinfo
  {author} {\bibfnamefont {R.~A.}\ \bibnamefont {Duine}},\ }\bibfield  {title}
  {\bibinfo {title} {New perspectives for {R}ashba spin–orbit coupling},\
  }\href {https://doi.org/10.1038/nmat4360} {\bibfield  {journal} {\bibinfo
  {journal} {Nat. Mater.}\ }\textbf {\bibinfo {volume} {14}},\ \bibinfo {pages}
  {871} (\bibinfo {year} {2015})}\BibitemShut {NoStop}%
\bibitem [{\citenamefont {Li}\ \emph {et~al.}(2017)\citenamefont {Li},
  \citenamefont {Lee}, \citenamefont {Huang}, \citenamefont {Burchesky},
  \citenamefont {Shteynas}, \citenamefont {Top}, \citenamefont {Jamison},\ and\
  \citenamefont {Ketterle}}]{LiJR}%
  \BibitemOpen
  \bibfield  {author} {\bibinfo {author} {\bibfnamefont {J.-R.}\ \bibnamefont
  {Li}}, \bibinfo {author} {\bibfnamefont {J.}~\bibnamefont {Lee}}, \bibinfo
  {author} {\bibfnamefont {W.}~\bibnamefont {Huang}}, \bibinfo {author}
  {\bibfnamefont {S.}~\bibnamefont {Burchesky}}, \bibinfo {author}
  {\bibfnamefont {B.}~\bibnamefont {Shteynas}}, \bibinfo {author}
  {\bibfnamefont {F.~{\c{C}}.}\ \bibnamefont {Top}}, \bibinfo {author}
  {\bibfnamefont {A.~O.}\ \bibnamefont {Jamison}},\ and\ \bibinfo {author}
  {\bibfnamefont {W.}~\bibnamefont {Ketterle}},\ }\bibfield  {title} {\bibinfo
  {title} {A stripe phase with supersolid properties in spin–orbit-coupled
  {B}ose–{E}instein condensates},\ }\href
  {https://doi.org/10.1038/nature21431} {\bibfield  {journal} {\bibinfo
  {journal} {Nature (London)}\ }\textbf {\bibinfo {volume} {543}},\ \bibinfo
  {pages} {91} (\bibinfo {year} {2017})}\BibitemShut {NoStop}%
\bibitem [{\citenamefont {Khamehchi}\ \emph {et~al.}(2017)\citenamefont
  {Khamehchi}, \citenamefont {Hossain}, \citenamefont {Mossman}, \citenamefont
  {Zhang}, \citenamefont {Busch}, \citenamefont {Forbes},\ and\ \citenamefont
  {Engels}}]{Khamehchi}%
  \BibitemOpen
  \bibfield  {author} {\bibinfo {author} {\bibfnamefont {M.~A.}\ \bibnamefont
  {Khamehchi}}, \bibinfo {author} {\bibfnamefont {K.}~\bibnamefont {Hossain}},
  \bibinfo {author} {\bibfnamefont {M.~E.}\ \bibnamefont {Mossman}}, \bibinfo
  {author} {\bibfnamefont {Y.}~\bibnamefont {Zhang}}, \bibinfo {author}
  {\bibfnamefont {T.}~\bibnamefont {Busch}}, \bibinfo {author} {\bibfnamefont
  {M.~M.}\ \bibnamefont {Forbes}},\ and\ \bibinfo {author} {\bibfnamefont
  {P.}~\bibnamefont {Engels}},\ }\bibfield  {title} {\bibinfo {title}
  {Negative-{M}ass {H}ydrodynamics in a {S}pin-{O}rbit--{C}oupled
  {B}ose-{E}instein {C}ondensate},\ }\href
  {https://doi.org/10.1103/PhysRevLett.118.155301} {\bibfield  {journal}
  {\bibinfo  {journal} {Phys. Rev. Lett.}\ }\textbf {\bibinfo {volume} {118}},\
  \bibinfo {pages} {155301} (\bibinfo {year} {2017})}\BibitemShut {NoStop}%
\bibitem [{\citenamefont {Kartashov}\ and\ \citenamefont
  {Zezyulin}(2019)}]{Kartashov}%
  \BibitemOpen
  \bibfield  {author} {\bibinfo {author} {\bibfnamefont {Y.~V.}\ \bibnamefont
  {Kartashov}}\ and\ \bibinfo {author} {\bibfnamefont {D.~A.}\ \bibnamefont
  {Zezyulin}},\ }\bibfield  {title} {\bibinfo {title} {Stable {M}ultiring and
  {R}otating {S}olitons in {T}wo-{D}imensional {S}pin-{O}rbit-{C}oupled
  {B}ose-{E}instein {C}ondensates with a {R}adially {P}eriodic {P}otential},\
  }\href {https://doi.org/10.1103/PhysRevLett.122.123201} {\bibfield  {journal}
  {\bibinfo  {journal} {Phys. Rev. Lett.}\ }\textbf {\bibinfo {volume} {122}},\
  \bibinfo {pages} {123201} (\bibinfo {year} {2019})}\BibitemShut {NoStop}%
\bibitem [{\citenamefont {Hasan}\ \emph {et~al.}(2022)\citenamefont {Hasan},
  \citenamefont {Madasu}, \citenamefont {Rathod}, \citenamefont {Kwong},
  \citenamefont {Miniatura}, \citenamefont {Chevy},\ and\ \citenamefont
  {Wilkowski}}]{Hasan}%
  \BibitemOpen
  \bibfield  {author} {\bibinfo {author} {\bibfnamefont {M.}~\bibnamefont
  {Hasan}}, \bibinfo {author} {\bibfnamefont {C.~S.}\ \bibnamefont {Madasu}},
  \bibinfo {author} {\bibfnamefont {K.~D.}\ \bibnamefont {Rathod}}, \bibinfo
  {author} {\bibfnamefont {C.~C.}\ \bibnamefont {Kwong}}, \bibinfo {author}
  {\bibfnamefont {C.}~\bibnamefont {Miniatura}}, \bibinfo {author}
  {\bibfnamefont {F.}~\bibnamefont {Chevy}},\ and\ \bibinfo {author}
  {\bibfnamefont {D.}~\bibnamefont {Wilkowski}},\ }\bibfield  {title} {\bibinfo
  {title} {Wave {Packet} {Dynamics} in {Synthetic} {Non}--{Abelian} {Gauge}
  {Fields}},\ }\href {https://doi.org/10.1103/PhysRevLett.129.130402}
  {\bibfield  {journal} {\bibinfo  {journal} {Phys. Rev. Lett.}\ }\textbf
  {\bibinfo {volume} {129}},\ \bibinfo {pages} {130402} (\bibinfo {year}
  {2022})}\BibitemShut {NoStop}%
\bibitem [{\citenamefont {Fr\"olian}\ \emph {et~al.}(2022)\citenamefont
  {Fr\"olian}, \citenamefont {Chisholm}, \citenamefont {Neri}, \citenamefont
  {Cabrera}, \citenamefont {Ramos}, \citenamefont {Celi},\ and\ \citenamefont
  {Tarruell}}]{Frolian}%
  \BibitemOpen
  \bibfield  {author} {\bibinfo {author} {\bibfnamefont {A.}~\bibnamefont
  {Fr\"olian}}, \bibinfo {author} {\bibfnamefont {C.~S.}\ \bibnamefont
  {Chisholm}}, \bibinfo {author} {\bibfnamefont {E.}~\bibnamefont {Neri}},
  \bibinfo {author} {\bibfnamefont {C.~R.}\ \bibnamefont {Cabrera}}, \bibinfo
  {author} {\bibfnamefont {R.}~\bibnamefont {Ramos}}, \bibinfo {author}
  {\bibfnamefont {A.}~\bibnamefont {Celi}},\ and\ \bibinfo {author}
  {\bibfnamefont {L.}~\bibnamefont {Tarruell}},\ }\bibfield  {title} {\bibinfo
  {title} {Realizing a 1{D} topological gauge theory in an optically dressed
  {B}{E}{C}},\ }\href {https://doi.org/10.1038/s41586-022-04943-3} {\bibfield
  {journal} {\bibinfo  {journal} {Nature (London)}\ }\textbf {\bibinfo {volume}
  {608}},\ \bibinfo {pages} {293} (\bibinfo {year} {2022})}\BibitemShut
  {NoStop}%
\bibitem [{\citenamefont {Wang}\ \emph {et~al.}(2010)\citenamefont {Wang},
  \citenamefont {Gao}, \citenamefont {Jian},\ and\ \citenamefont
  {Zhai}}]{Wang}%
  \BibitemOpen
  \bibfield  {author} {\bibinfo {author} {\bibfnamefont {C.}~\bibnamefont
  {Wang}}, \bibinfo {author} {\bibfnamefont {C.}~\bibnamefont {Gao}}, \bibinfo
  {author} {\bibfnamefont {C.-M.}\ \bibnamefont {Jian}},\ and\ \bibinfo
  {author} {\bibfnamefont {H.}~\bibnamefont {Zhai}},\ }\bibfield  {title}
  {\bibinfo {title} {Spin-{Orbit} {Coupled} {Spinor} {Bose}-{Einstein}
  {Condensates}},\ }\href {https://doi.org/10.1103/PhysRevLett.105.160403}
  {\bibfield  {journal} {\bibinfo  {journal} {Phys. Rev. Lett.}\ }\textbf
  {\bibinfo {volume} {105}},\ \bibinfo {pages} {160403} (\bibinfo {year}
  {2010})}\BibitemShut {NoStop}%
\bibitem [{\citenamefont {Ho}\ and\ \citenamefont {Zhang}(2011)}]{Ho2011}%
  \BibitemOpen
  \bibfield  {author} {\bibinfo {author} {\bibfnamefont {T.-L.}\ \bibnamefont
  {Ho}}\ and\ \bibinfo {author} {\bibfnamefont {S.}~\bibnamefont {Zhang}},\
  }\bibfield  {title} {\bibinfo {title} {Bose-{E}instein {C}ondensates with
  {S}pin-{O}rbit interaction},\ }\href
  {https://doi.org/10.1103/PhysRevLett.107.150403} {\bibfield  {journal}
  {\bibinfo  {journal} {Phys. Rev. Lett.}\ }\textbf {\bibinfo {volume} {107}},\
  \bibinfo {pages} {150403} (\bibinfo {year} {2011})}\BibitemShut {NoStop}%
\bibitem [{\citenamefont {Hu}\ \emph {et~al.}(2012)\citenamefont {Hu},
  \citenamefont {Ramachandhran}, \citenamefont {Pu},\ and\ \citenamefont
  {Liu}}]{Hu2012}%
  \BibitemOpen
  \bibfield  {author} {\bibinfo {author} {\bibfnamefont {H.}~\bibnamefont
  {Hu}}, \bibinfo {author} {\bibfnamefont {B.}~\bibnamefont {Ramachandhran}},
  \bibinfo {author} {\bibfnamefont {H.}~\bibnamefont {Pu}},\ and\ \bibinfo
  {author} {\bibfnamefont {X.-J.}\ \bibnamefont {Liu}},\ }\bibfield  {title}
  {\bibinfo {title} {Spin-{Orbit} {Coupled} {Weakly} {Interacting}
  {Bose}-{Einstein} {Condensates} in {Harmonic} {Traps}},\ }\href
  {https://doi.org/10.1103/PhysRevLett.108.010402} {\bibfield  {journal}
  {\bibinfo  {journal} {Phys. Rev. Lett.}\ }\textbf {\bibinfo {volume} {108}},\
  \bibinfo {pages} {010402} (\bibinfo {year} {2012})}\BibitemShut {NoStop}%
\bibitem [{\citenamefont {Zhang}\ \emph {et~al.}(2012)\citenamefont {Zhang},
  \citenamefont {Mao},\ and\ \citenamefont {Zhang}}]{Zhang2012}%
  \BibitemOpen
  \bibfield  {author} {\bibinfo {author} {\bibfnamefont {Y.}~\bibnamefont
  {Zhang}}, \bibinfo {author} {\bibfnamefont {L.}~\bibnamefont {Mao}},\ and\
  \bibinfo {author} {\bibfnamefont {C.}~\bibnamefont {Zhang}},\ }\bibfield
  {title} {\bibinfo {title} {Mean-{Field} {Dynamics} of {Spin}-{Orbit}
  {Coupled} {Bose}-{Einstein} {Condensates}},\ }\href
  {https://doi.org/10.1103/PhysRevLett.108.035302} {\bibfield  {journal}
  {\bibinfo  {journal} {Phys. Rev. Lett.}\ }\textbf {\bibinfo {volume} {108}},\
  \bibinfo {pages} {035302} (\bibinfo {year} {2012})}\BibitemShut {NoStop}%
\bibitem [{\citenamefont {Song}\ \emph {et~al.}(2013)\citenamefont {Song},
  \citenamefont {Zhang}, \citenamefont {Wen},\ and\ \citenamefont
  {Wang}}]{Song2013}%
  \BibitemOpen
  \bibfield  {author} {\bibinfo {author} {\bibfnamefont {S.-W.}\ \bibnamefont
  {Song}}, \bibinfo {author} {\bibfnamefont {Y.-C.}\ \bibnamefont {Zhang}},
  \bibinfo {author} {\bibfnamefont {L.}~\bibnamefont {Wen}},\ and\ \bibinfo
  {author} {\bibfnamefont {H.}~\bibnamefont {Wang}},\ }\bibfield  {title}
  {\bibinfo {title} {Spin–orbit coupling induced displacement and hidden spin
  textures in spin-1 {B}ose–{E}instein condensates},\ }\href
  {https://doi.org/10.1088/0953-4075/46/14/145304} {\bibfield  {journal}
  {\bibinfo  {journal} {J. Phys. B}\ }\textbf {\bibinfo {volume} {46}},\
  \bibinfo {pages} {145304} (\bibinfo {year} {2013})}\BibitemShut {NoStop}%
\bibitem [{\citenamefont {Sakaguchi}\ \emph {et~al.}(2014)\citenamefont
  {Sakaguchi}, \citenamefont {Li},\ and\ \citenamefont {Malomed}}]{Sakaguchi}%
  \BibitemOpen
  \bibfield  {author} {\bibinfo {author} {\bibfnamefont {H.}~\bibnamefont
  {Sakaguchi}}, \bibinfo {author} {\bibfnamefont {B.}~\bibnamefont {Li}},\ and\
  \bibinfo {author} {\bibfnamefont {B.~A.}\ \bibnamefont {Malomed}},\
  }\bibfield  {title} {\bibinfo {title} {Creation of two-dimensional composite
  solitons in spin-orbit-coupled self-attractive {B}ose-{E}instein condensates
  in free space},\ }\href {https://doi.org/10.1103/PhysRevE.89.032920}
  {\bibfield  {journal} {\bibinfo  {journal} {Phys. Rev. E}\ }\textbf {\bibinfo
  {volume} {89}},\ \bibinfo {pages} {032920} (\bibinfo {year}
  {2014})}\BibitemShut {NoStop}%
\bibitem [{\citenamefont {Xu}\ \emph {et~al.}(2015)\citenamefont {Xu},
  \citenamefont {Zhang},\ and\ \citenamefont {Zhang}}]{Xu2015}%
  \BibitemOpen
  \bibfield  {author} {\bibinfo {author} {\bibfnamefont {Y.}~\bibnamefont
  {Xu}}, \bibinfo {author} {\bibfnamefont {Y.}~\bibnamefont {Zhang}},\ and\
  \bibinfo {author} {\bibfnamefont {C.}~\bibnamefont {Zhang}},\ }\bibfield
  {title} {\bibinfo {title} {Bright solitons in a two-dimensional
  spin-orbit-coupled dipolar {B}ose-{E}instein condensate},\ }\href
  {https://doi.org/10.1103/PhysRevA.92.013633} {\bibfield  {journal} {\bibinfo
  {journal} {Phys. Rev. A}\ }\textbf {\bibinfo {volume} {92}},\ \bibinfo
  {pages} {013633} (\bibinfo {year} {2015})}\BibitemShut {NoStop}%
\bibitem [{\citenamefont {Wang}\ \emph {et~al.}(2020)\citenamefont {Wang},
  \citenamefont {Wen}, \citenamefont {Chen}, \citenamefont {Zhang},\ and\
  \citenamefont {Zhang}}]{Wang2020}%
  \BibitemOpen
  \bibfield  {author} {\bibinfo {author} {\bibfnamefont {Y.-J.}\ \bibnamefont
  {Wang}}, \bibinfo {author} {\bibfnamefont {L.}~\bibnamefont {Wen}}, \bibinfo
  {author} {\bibfnamefont {G.-P.}\ \bibnamefont {Chen}}, \bibinfo {author}
  {\bibfnamefont {S.-G.}\ \bibnamefont {Zhang}},\ and\ \bibinfo {author}
  {\bibfnamefont {X.-F.}\ \bibnamefont {Zhang}},\ }\bibfield  {title} {\bibinfo
  {title} {Formation, stability, and dynamics of vector bright solitons in a
  trapless {B}ose–{E}instein condensate with spin–orbit coupling},\ }\href
  {https://doi.org/10.1088/1367-2630/ab725b} {\bibfield  {journal} {\bibinfo
  {journal} {New J. Phys.}\ }\textbf {\bibinfo {volume} {22}},\ \bibinfo
  {pages} {033006} (\bibinfo {year} {2020})}\BibitemShut {NoStop}%
\bibitem [{\citenamefont {Kato}\ \emph {et~al.}(2017)\citenamefont {Kato},
  \citenamefont {Zhang},\ and\ \citenamefont {Saito}}]{Kato2017}%
  \BibitemOpen
  \bibfield  {author} {\bibinfo {author} {\bibfnamefont {M.}~\bibnamefont
  {Kato}}, \bibinfo {author} {\bibfnamefont {X.-F.}\ \bibnamefont {Zhang}},\
  and\ \bibinfo {author} {\bibfnamefont {H.}~\bibnamefont {Saito}},\ }\bibfield
   {title} {\bibinfo {title} {Vortex pairs in a spin-orbit-coupled
  {B}ose-{E}instein condensate},\ }\href
  {https://doi.org/10.1103/PhysRevA.95.043605} {\bibfield  {journal} {\bibinfo
  {journal} {Phys. Rev. A}\ }\textbf {\bibinfo {volume} {95}},\ \bibinfo
  {pages} {043605} (\bibinfo {year} {2017})}\BibitemShut {NoStop}%
\bibitem [{\citenamefont {Ravisankar}\ \emph {et~al.}(2021)\citenamefont
  {Ravisankar}, \citenamefont {Fabrelli}, \citenamefont {Gammal}, \citenamefont
  {Muruganandam},\ and\ \citenamefont {Mishra}}]{Ravisankar2021}%
  \BibitemOpen
  \bibfield  {author} {\bibinfo {author} {\bibfnamefont {R.}~\bibnamefont
  {Ravisankar}}, \bibinfo {author} {\bibfnamefont {H.}~\bibnamefont
  {Fabrelli}}, \bibinfo {author} {\bibfnamefont {A.}~\bibnamefont {Gammal}},
  \bibinfo {author} {\bibfnamefont {P.}~\bibnamefont {Muruganandam}},\ and\
  \bibinfo {author} {\bibfnamefont {P.~K.}\ \bibnamefont {Mishra}},\ }\bibfield
   {title} {\bibinfo {title} {Effect of {R}ashba spin-orbit and {R}abi
  couplings on the excitation spectrum of binary {B}ose-{E}instein
  condensates},\ }\href {https://doi.org/10.1103/PhysRevA.104.053315}
  {\bibfield  {journal} {\bibinfo  {journal} {Phys. Rev. A}\ }\textbf {\bibinfo
  {volume} {104}},\ \bibinfo {pages} {053315} (\bibinfo {year}
  {2021})}\BibitemShut {NoStop}%
\bibitem [{\citenamefont {Mardonov}\ \emph
  {et~al.}(2015{\natexlab{a}})\citenamefont {Mardonov}, \citenamefont
  {Modugno},\ and\ \citenamefont {Sherman}}]{MardonovPRL2015}%
  \BibitemOpen
  \bibfield  {author} {\bibinfo {author} {\bibfnamefont {S.}~\bibnamefont
  {Mardonov}}, \bibinfo {author} {\bibfnamefont {M.}~\bibnamefont {Modugno}},\
  and\ \bibinfo {author} {\bibfnamefont {E.~Y.}\ \bibnamefont {Sherman}},\
  }\bibfield  {title} {\bibinfo {title} {Dynamics of spin-orbit coupled
  bose-einstein condensates in a random potential},\ }\href
  {https://doi.org/10.1103/PhysRevLett.115.180402} {\bibfield  {journal}
  {\bibinfo  {journal} {Phys. Rev. Lett.}\ }\textbf {\bibinfo {volume} {115}},\
  \bibinfo {pages} {180402} (\bibinfo {year} {2015}{\natexlab{a}})}\BibitemShut
  {NoStop}%
\bibitem [{\citenamefont {Mardonov}\ \emph
  {et~al.}(2015{\natexlab{b}})\citenamefont {Mardonov}, \citenamefont
  {Sherman}, \citenamefont {Muga}, \citenamefont {Wang}, \citenamefont {Ban},\
  and\ \citenamefont {Chen}}]{MardonovPRA2015}%
  \BibitemOpen
  \bibfield  {author} {\bibinfo {author} {\bibfnamefont {S.}~\bibnamefont
  {Mardonov}}, \bibinfo {author} {\bibfnamefont {E.~Y.}\ \bibnamefont
  {Sherman}}, \bibinfo {author} {\bibfnamefont {J.~G.}\ \bibnamefont {Muga}},
  \bibinfo {author} {\bibfnamefont {H.-W.}\ \bibnamefont {Wang}}, \bibinfo
  {author} {\bibfnamefont {Y.}~\bibnamefont {Ban}},\ and\ \bibinfo {author}
  {\bibfnamefont {X.}~\bibnamefont {Chen}},\ }\bibfield  {title} {\bibinfo
  {title} {Collapse of spin-orbit-coupled bose-einstein condensates},\ }\href
  {https://doi.org/10.1103/PhysRevA.91.043604} {\bibfield  {journal} {\bibinfo
  {journal} {Phys. Rev. A}\ }\textbf {\bibinfo {volume} {91}},\ \bibinfo
  {pages} {043604} (\bibinfo {year} {2015}{\natexlab{b}})}\BibitemShut
  {NoStop}%
\bibitem [{\citenamefont {Mardonov}\ \emph {et~al.}(2018)\citenamefont
  {Mardonov}, \citenamefont {Konotop}, \citenamefont {Malomed}, \citenamefont
  {Modugno},\ and\ \citenamefont {Sherman}}]{MardonovPRA2018}%
  \BibitemOpen
  \bibfield  {author} {\bibinfo {author} {\bibfnamefont {S.}~\bibnamefont
  {Mardonov}}, \bibinfo {author} {\bibfnamefont {V.~V.}\ \bibnamefont
  {Konotop}}, \bibinfo {author} {\bibfnamefont {B.~A.}\ \bibnamefont
  {Malomed}}, \bibinfo {author} {\bibfnamefont {M.}~\bibnamefont {Modugno}},\
  and\ \bibinfo {author} {\bibfnamefont {E.~Y.}\ \bibnamefont {Sherman}},\
  }\bibfield  {title} {\bibinfo {title} {Spin-orbit-coupled soliton in a random
  potential},\ }\href {https://doi.org/10.1103/PhysRevA.98.023604} {\bibfield
  {journal} {\bibinfo  {journal} {Phys. Rev. A}\ }\textbf {\bibinfo {volume}
  {98}},\ \bibinfo {pages} {023604} (\bibinfo {year} {2018})}\BibitemShut
  {NoStop}%
\bibitem [{\citenamefont {Ozawa}\ and\ \citenamefont {Baym}(2012)}]{Ozawa}%
  \BibitemOpen
  \bibfield  {author} {\bibinfo {author} {\bibfnamefont {T.}~\bibnamefont
  {Ozawa}}\ and\ \bibinfo {author} {\bibfnamefont {G.}~\bibnamefont {Baym}},\
  }\bibfield  {title} {\bibinfo {title} {Ground-state phases of ultracold
  bosons with {Rashba}-{Dresselhaus} spin-orbit coupling},\ }\href
  {https://doi.org/10.1103/PhysRevA.85.013612} {\bibfield  {journal} {\bibinfo
  {journal} {Phys. Rev. A}\ }\textbf {\bibinfo {volume} {85}},\ \bibinfo
  {pages} {013612} (\bibinfo {year} {2012})}\BibitemShut {NoStop}%
\bibitem [{\citenamefont {Gautam}\ and\ \citenamefont
  {Adhikari}(2014)}]{Gautam}%
  \BibitemOpen
  \bibfield  {author} {\bibinfo {author} {\bibfnamefont {S.}~\bibnamefont
  {Gautam}}\ and\ \bibinfo {author} {\bibfnamefont {S.~K.}\ \bibnamefont
  {Adhikari}},\ }\bibfield  {title} {\bibinfo {title} {Phase separation in a
  spin-orbit-coupled {Bose}-{Einstein} condensate},\ }\href
  {https://doi.org/10.1103/PhysRevA.90.043619} {\bibfield  {journal} {\bibinfo
  {journal} {Phys. Rev. A}\ }\textbf {\bibinfo {volume} {90}},\ \bibinfo
  {pages} {043619} (\bibinfo {year} {2014})}\BibitemShut {NoStop}%
\bibitem [{\citenamefont {Merhasin}\ \emph {et~al.}(2005)\citenamefont
  {Merhasin}, \citenamefont {Malomed},\ and\ \citenamefont
  {Driben}}]{Merhasin}%
  \BibitemOpen
  \bibfield  {author} {\bibinfo {author} {\bibfnamefont {I.~M.}\ \bibnamefont
  {Merhasin}}, \bibinfo {author} {\bibfnamefont {B.~A.}\ \bibnamefont
  {Malomed}},\ and\ \bibinfo {author} {\bibfnamefont {R.}~\bibnamefont
  {Driben}},\ }\bibfield  {title} {\bibinfo {title} {Transition to miscibility
  in a binary {B}ose–{E}instein condensate induced by linear coupling},\
  }\href {https://doi.org/10.1088/0953-4075/38/7/009} {\bibfield  {journal}
  {\bibinfo  {journal} {J. Phys. B}\ }\textbf {\bibinfo {volume} {38}},\
  \bibinfo {pages} {877} (\bibinfo {year} {2005})}\BibitemShut {NoStop}%
\bibitem [{\citenamefont {Beeler}\ \emph {et~al.}(2013)\citenamefont {Beeler},
  \citenamefont {Williams}, \citenamefont {Jim\'enez-Garc\'ia}, \citenamefont
  {LeBlanc}, \citenamefont {Perry},\ and\ \citenamefont {Spielman}}]{Beeler}%
  \BibitemOpen
  \bibfield  {author} {\bibinfo {author} {\bibfnamefont {M.~C.}\ \bibnamefont
  {Beeler}}, \bibinfo {author} {\bibfnamefont {R.~A.}\ \bibnamefont
  {Williams}}, \bibinfo {author} {\bibfnamefont {K.}~\bibnamefont
  {Jim\'enez-Garc\'ia}}, \bibinfo {author} {\bibfnamefont {L.~J.}\ \bibnamefont
  {LeBlanc}}, \bibinfo {author} {\bibfnamefont {A.~R.}\ \bibnamefont {Perry}},\
  and\ \bibinfo {author} {\bibfnamefont {I.~B.}\ \bibnamefont {Spielman}},\
  }\bibfield  {title} {\bibinfo {title} {The spin {H}all effect in a quantum
  gas},\ }\href {https://doi.org/10.1038/nature12185} {\bibfield  {journal}
  {\bibinfo  {journal} {Nature (London)}\ }\textbf {\bibinfo {volume} {498}},\
  \bibinfo {pages} {201} (\bibinfo {year} {2013})}\BibitemShut {NoStop}%
\bibitem [{\citenamefont {Ramachandhran}\ \emph {et~al.}(2012)\citenamefont
  {Ramachandhran}, \citenamefont {Opanchuk}, \citenamefont {Liu}, \citenamefont
  {Pu}, \citenamefont {Drummond},\ and\ \citenamefont {Hu}}]{Ramachandhran}%
  \BibitemOpen
  \bibfield  {author} {\bibinfo {author} {\bibfnamefont {B.}~\bibnamefont
  {Ramachandhran}}, \bibinfo {author} {\bibfnamefont {B.}~\bibnamefont
  {Opanchuk}}, \bibinfo {author} {\bibfnamefont {X.-J.}\ \bibnamefont {Liu}},
  \bibinfo {author} {\bibfnamefont {H.}~\bibnamefont {Pu}}, \bibinfo {author}
  {\bibfnamefont {P.~D.}\ \bibnamefont {Drummond}},\ and\ \bibinfo {author}
  {\bibfnamefont {H.}~\bibnamefont {Hu}},\ }\bibfield  {title} {\bibinfo
  {title} {Half-quantum vortex state in a spin-orbit-coupled {Bose}-{Einstein}
  condensate},\ }\href {https://doi.org/10.1103/PhysRevA.85.023606} {\bibfield
  {journal} {\bibinfo  {journal} {Phys. Rev. A}\ }\textbf {\bibinfo {volume}
  {85}},\ \bibinfo {pages} {023606} (\bibinfo {year} {2012})}\BibitemShut
  {NoStop}%
\end{thebibliography}%

\end{document}